\documentclass[%
twocolumn,
superscriptaddress,
nofootinbib,
 amsmath,amssymb,
 aps,
]{revtex4-2}

\usepackage{microtype}
\usepackage{graphicx}
\usepackage{dcolumn}
\usepackage{bm}

\usepackage{amssymb}
\usepackage{amsmath}
\usepackage{amsthm}
\usepackage{mathtools}
\usepackage[usenames,dvipsnames]{xcolor}
\usepackage{epsfig}
\usepackage{dcolumn}
\usepackage{tikz}
\usepackage{tikz-cd}
\usetikzlibrary{calc}
\usepackage{braket}
\usepackage{adjustbox}
\usepackage{hyperref}
\usepackage{verbatim}
\hypersetup{
	colorlinks=true,
	urlcolor=Maroon,
	linkcolor=RoyalBlue,
	citecolor=Maroon,
	bookmarks=true,
	pdftitle={Feynman Polytopes and the Tropical Geometry of UV and IR Divergences},
	pdfauthor={Nima Arkani-Hamed, Aaron Hillman, Sebastian Mizera},
	pdfdisplaydoctitle=true,
	pdfstartview=FitH
}
\definecolor{color2}{rgb}{0.368417, 0.506779, 0.709798}
\definecolor{color3}{rgb}{0.880722, 0.611041, 0.142051}
\definecolor{color5}{rgb}{0.560181, 0.691569, 0.194885}
\definecolor{color1}{rgb}{0.922526, 0.385626, 0.209179}
\definecolor{color6}{rgb}{0.528488, 0.470624, 0.701351}
\definecolor{color4}{rgb}{0.772079, 0.431554, 0.102387}

\usetikzlibrary{decorations.pathmorphing}
\tikzset{massless/.style={solid, line width=1.1pt}}
\tikzset{massive/.style={solid, line width=1.25pt, double}}
\tikzset{massive2/.style={solid, line width=0.9pt, double}}

\def \U {\mathcal{U}}
\def \F {\mathcal{F}}
\def \I {\mathcal{I}}
\def \J {\mathcal{J}}
\def \UU {\mathbf{U}}
\def \FF {\mathbf{F}}
\def \SS {\mathbf{S}}
\def \LL {\mathrm{L}}
\def \L {\mathrm{L}}
\def \E {\mathrm{E}}
\def \D {\mathrm{D}}
\def \d {\mathrm{d}}
\def \leq {\leqslant}
\def \geq {\geqslant}
\def \Newt {\mathrm{Newt}}
\def \nn {\nonumber}
\def \Trop {\mathrm{Trop}}

\begin{document}

\title{Feynman Polytopes and the Tropical Geometry of UV and IR Divergences}

\author{Nima Arkani-Hamed}
\affiliation{%
Institute for Advanced Study, Princeton, NJ 08540, USA
}
\author{Aaron Hillman}
\affiliation{%
Department of Physics, Jadwin Hall, Princeton University, NJ 08540, USA
}%
\author{Sebastian Mizera}%
\affiliation{%
Institute for Advanced Study, Princeton, NJ 08540, USA
}

\begin{abstract}
We introduce a class of polytopes that concisely capture the structure of UV and IR divergences of general Feynman integrals in Schwinger parameter space, treating them in a unified way as worldline segments shrinking and expanding at different relative rates. While these polytopes conventionally arise as convex hulls---via Newton polytopes of Symanzik polynomials---we show that they also have a remarkably simple dual description as cut out by linear inequalities defining the facets. It is this dual definition that makes it possible to transparently understand and efficiently compute leading UV and IR divergences for any Feynman integral. In the case of the UV, this provides a transparent geometric understanding of the familiar nested and overlapping divergences. In the IR, the polytope exposes a new perspective on soft/collinear singularities and their intricate generalizations. Tropical geometry furnishes a simple framework for calculating the leading UV/IR divergences of any Feynman integral, associating them with the volumes of certain dual cones. As concrete applications, we generalize  Weinberg's theorem to include a characterization of IR divergences, and  classify space-time dimensions in which general IR divergences (logarithmic as well as power-law) can occur. We also compute the leading IR divergence of rectangular fishnet diagrams at all loop orders, which turn out to have a surprisingly simple combinatorial description.
\end{abstract}

\maketitle
\section{Introduction}

One of the key outstanding questions in the modern S-matrix program is understanding the renormalization group (RG) from an on-shell point of view. The Wilsonian picture is fundamentally Euclidean, and referring to ``coarse graining" in position space, or ``integrating out high momentum modes" in momentum space, crucially relies on concepts far removed from the on-shell philosophy. Alternatively, much of the technical content of the RG was earlier realized using the seemingly more complicated ideas of  Bogoliubov--Parasiuk--Hepp--Zimmermann \cite{10.1007/BF02392399,Hepp1966,Zimmermann1969} on ultraviolet (UV) divergences, which have been given a deeper Hopf-algebraic understanding \cite{Connes:1998qv,Connes:1999yr,Connes:2000fe} exhibiting a closer connection to on-shell ideas.  More recently,  analogous RG-like structures were found for infrared (IR) divergences in the context of QCD \cite{Mueller:1979ih,Collins:1980ih,Sen:1981sd,Korchemsky:1988si,Sterman:2002qn,Gardi:2009qi,Agarwal:2021ais} and later soft-collinear effective theory, see, e.g., \cite{Collins:1989gx,Becher:2009qa,StewartLectures,Becher:2014oda,Ma:2019hjq,Beekveldt:2020kzk}.

It is then natural to ask whether these perspectives on UV and IR divergences can be unified into a single picture, one which can connect the emerging geometric underpinnings of scattering amplitudes to the physics of UV divergences and the RG. In this note, we take a step in this direction, by describing a beautiful class of polytopes associated with any Feynman graph, whose facet structure concisely captures and calculates both the leading UV and IR divergences in a unified way. 

Our starting point is the treatment of Feynman integrals in dimensional regularization from a worldline perspective, where instead of integrating over the loop momenta, the only variables are the Schwinger proper times $\alpha=e^\tau$ measuring lengths of each propagator. From this point of view, the limits $\tau \to \mp\infty$ correspond to shrinking and expanding edges, which are naturally associated to the UV and IR singularities respectively. They are thus put on the same footing and the unified analysis of all divergences boils down to examining the asymptotic behavior of the integrand at infinity. Mathematicians have studied this subject, known as \emph{tropical geometry}, for decades, see, e.g., \cite{maclagan2015introduction} for an introduction. We will explore the physical meaning and consequences of this geometry in our setting.

\begin{figure}
    \centering
    \includegraphics[]{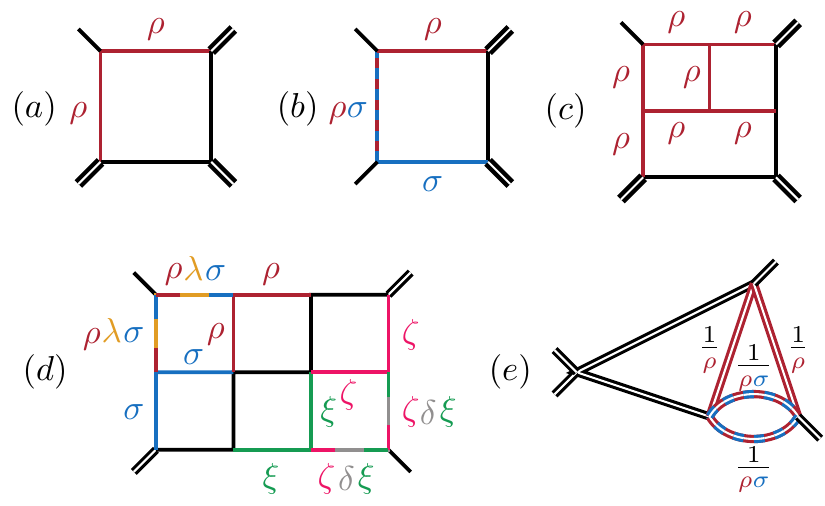}
    \caption{\label{fig:scalings}Example scalings of Schwinger parameter responsible for: (a) collinear, (b) soft/collinear, (c) power-law IR, (d) nested soft/collinear, and (e) nested UV singularities in four dimensions. In each case the divergence comes from shrinking/expanding the relevant worldline edge lengths as the product of the variables $\rho,\sigma, \ldots$ to the indicated powers, as they all scale  $ \to \infty$.  Massless and massive particles are denoted with single and doubled lines respectively.}
\end{figure}

As we will see, all of the elaborate structure of UV and IR divergences of a Feynman integral associated with a graph $G$ is captured by the \emph{Feynman polytope} $\SS_G$. It can be constructed out of the \emph{Symanzik polytopes} (previously explored in \cite{brown2011multiple,Brown:2015fyf,Schultka:2018nrs,Borinsky:2020rqs,Mizera:2021icv} for IR-safe kinematics) defined fully by the combinatorics of $G$ and the kinematic data of external momenta and particle masses. 

Elements of polyhedral geometry of Feynman integrals previously appeared in sector decomposition \cite{Binoth:2000ps,Binoth:2003ak,Bogner:2007cr,Kaneko:2009qx,Kaneko:2010kj,Borowka:2015mxa,Schlenk:2016epj,Heinrich:2021dbf}, expansion by regions \cite{Pak:2010pt,Ananthanarayan:2018tog,Semenova:2018cwy,Ananthanarayan:2020fhl,Tellander:2021xdz}, and blowups of integration domains \cite{Bloch2006,Brown:2015fyf}.\footnote{Tropical geometry has also recently emerged as a useful tool in studying other, complementary, aspects of scattering amplitudes, see, e.g., \cite{Tourkine:2013rda,Panzer:2019yxl,Cachazo:2019ngv,Drummond:2020kqg,Arkani-Hamed:2020cig,Drummond:2019cxm,Arkani-Hamed:2019mrd,Arkani-Hamed:2019plo,Arkani-Hamed:2020tuz,Chicherin:2020umh,Cachazo:2020wgu,He:2021non}.}
Sector decomposition is a powerful general algorithm for studying singularities of generalized hypergeometric function (see, e.g., \cite{Mastrolia:2018uzb,delaCruz:2019skx,Klausen:2019hrg,Feng:2019bdx,Abreu:2019wzk,Tellander:2021xdz,Mizera:2021icv} for recent progress).  But as such, its application to the study of Feynman diagrams does not make use of any special properties that one might expect would arise for the integrals which arise from physics, meaning one has to laboriously compute $\SS_G$ for every $G$ from scratch. One of our central results is that the structure of the integrals handed to us by physics, and the corresponding polyhedral geometry, is indeed very special, allowing us to give an analytic solution to this problem for any scalar Feynman diagram. 

The polytopes begin their lives described as the convex hull of the exponent vectors of Symanzik polynomials. We show that they also have a beautifully simple \emph{facet description}, cut out by linear inequalities. It is this dual definition that makes it possible to understand and compute the leading UV and IR divergences.

An overarching theme in the emerging connection between combinatorial geometry and scattering amplitudes has been the realization that the singularities associated with ``factorization" of amplitudes are reflected in an analogous ``factorization" seen in the facet structure of the geometries. This is also seen in the Symanzik polytopes and ${\bf S}_G$, whose boundary structure transparently encodes the combinatorics of shrinking subgraphs of $G$, thereby precisely encoding the ``Russian doll" hierarchical separation of worldline edge-lengths in $G$ that have the potential to generate divergences. 

The definition of ${\bf S}_G$ as cut out by inequalities further endows each face of $\SS_G$ with a number: its UV/IR degree of divergence. The way these faces fit into the polytope dictates how the individual divergences are related to one another. In the UV, this prescription recovers the standard picture of nested and overlapping divergences, in agreement with \cite{Brown:2011pj,Brown:2015fyf}. In the IR, it reveals novel patterns of divergences, including the familiar soft/collinear divergences together with their intricate generalizations. Examples are shown in Fig.~\ref{fig:scalings}.

As an extension of \cite{Arkani-Hamed:2019mrd}, we explain how the leading divergent term of every Feynman integral can be computed in terms of volumes of dual cones. In this formalism, the divergence is viewed as an additional redundancy of the integrand in certain directions at infinity. We give a prescription for computing the coefficient of this divergence as a procedure of ``modding out'' by the enhanced symmetry. We apply this tool to various examples, including the rectangular fishnet diagrams, resulting in a formula for the leading IR divergence at any loop order.

Our goal in this note is to take a geodesic path to the central constructions, and illustrate their utility and power with a few examples. A more leisurely description of these ideas as well as further generalizations and examples will appear in \cite{AHHM}. 

\section{Schwinger Parametrization}
A scalar Feynman integral in parameter space corresponding to a graph $G$ is given by 
\begin{equation}\label{eq:I}
    \I_G = \Gamma(d) \int \frac{\d^\E \alpha}{\text{GL}(1)} \frac{1}{\mathcal{U}^{\D/2-d}\mathcal{F}^{d}},
\end{equation}
where $d = \E-\L\D/2$ is the superficial degree of divergence for a scalar diagram $G$ with $\E$ edges and $\L$ loops in dimension $\D$.  The functions in the integral are the Symanzik polynomials which admit a graphical expansion in terms of spanning trees $T$ and $2$-trees $T'$
\begin{equation}
    \mathcal{U} = \sum_T \prod\limits_{e \notin T}\alpha_e, \quad
    \mathcal{F} = \sum_{T'} p_{T'}^2 \!\!\prod\limits_{e \notin T'}\alpha_e - \U \sum_{e=1}^{\E} m_e^2\alpha_e,
\end{equation}
where $p_{T'}^\mu$ is the total momentum flowing across the $2$-tree, see, e.g., \cite{nakanishi1971graph}. Modding out by overall-scale $\mathrm{GL}(1)$ redundancy means we can set $\alpha_\E=1$. Including Feynman integrals with non-trivial numerators does not pose difficulty and will be considered in \cite{AHHM}.

Below, we will use $\gamma$ to mean a subgraph of $G$, and $G/\gamma$ to denote $G$ with each connected component of $\gamma$ contracted to a vertex. To avoid confusion, we add the subscript ${}_G$ to denote the quantity for a specific diagram $G$, e.g., $\L_\gamma$ denotes the number of loops in $\gamma$.

\section{\label{sec:general}Tropical Geometry}

Feynman integrals \eqref{eq:I} fall into a general class of integrals whose properties can be understood using tropical geometry \cite{gelfand2009discriminants,Nilsson2013,berkesch2013eulermellin,Arkani-Hamed:2019mrd}. In terms of the variables $\tau_e = \log\alpha_e$, the integrand of \eqref{eq:I} can be approximated at infinity, where the UV/IR divergences come from, by $e^{\Trop}$ with the piecewise-linear function
\begin{equation}\label{tropdef}
\Trop = \tau_G + (d_G -\tfrac{\D}{2})\max(\U_G) - d_G \max (\F_G), 
\end{equation}
where $\tau_\gamma = \sum_{e \in \gamma}\tau_e$ and the maxima are taken over all the monomials in $\U_G$ and $\F_G$. Equivalently, one can capture the same information by the Minkowski sum/difference $\SS_G = \UU_G \oplus c \FF_G$, of the Newton polytopes
\begin{equation}\label{eq:UF}
    \UU_G = \Newt(\mathcal{U}_G), \qquad \FF_G = \Newt(\mathcal{F}_G),
\end{equation}
which we refer to as \textit{Symanzik polytopes} and study closely in Sec.~\ref{sec:Symanzik}.
To obtain the most refined information about the regions where divergences can possibly come from, it is sufficient to consider $c>0$, see App.~\ref{app:general}.
Every \emph{ray} (boundary of domains of linearity) of $\Trop$ defines the normal to a facet of the Feynman polytope $\SS_G$ and hence the two pictures can be used interchangeably.

\paragraph*{\bf Leading divergences.} Divergences come from cones (faces) generated by $\Trop \geq 0$ rays. We focus on logarithmic cases with $\Trop = 0$. Such $m$-dimensional cones (codimension-$m$ faces) give $\log^m$ or $1/\varepsilon^m$ UV/IR singularities. The leading divergence is controlled by cones $C_F$, or \emph{maximally-divergent faces} $F$ for which $m$ is the largest. As a refinement of the standard tropical rules \cite{Arkani-Hamed:2019mrd}, we find
\begin{equation}\label{tropvolume}
\I_G = \Gamma(d_G) \sum_{F} \mathrm{vol}(C_{F})\, \J_{F} + \ldots.
\end{equation}
It involves two ingredients: the volume of the cone $C_{F}$ (bounded by $\Trop \geq -1$) gives a numerical factor $\propto 1/\varepsilon^{m}$, while all the kinematic dependence is captured by the modded integral
\begin{equation}
\J_F =  \int \frac{\d^\E \alpha}{\text{GL}(1)^{m+1}}\, \frac{1}{\mathcal{U}_G^{\D/2-d}\mathcal{F}_G^{d}} \bigg|_{C_{F}^\perp}.
\end{equation}
Here we use the tropical approximation for the integrand in the direction $C_F^\perp$ orthogonal to the cone $C_F$, which by definition has an enhanced $\mathrm{GL}(1)^m$ symmetry. So extracting the coefficient of this divergence is precisely modding out by the extra redundancy!
Strictly speaking, the volume and the modded integral are not uniquely defined, but their product is, as illustrated on examples in App.~\ref{app:further-examples}.

\section{\label{sec:Symanzik}Symanzik Polytopes}

The polytopes $\UU_G$ and $\FF_G$ have been defined as a convex hull of vertices via the Newton polytopes \eqref{eq:UF}.  But the singularities of Feynman integrals are instead tied to their \emph{facet} description, cutting out the polytopes with linear inequalities. 

Now, there is a systematic algorithm for finding the facets of an $n$-dimensional polytope given its vertices: checking whether a collection of vertices is a facet by asking whether the remaining vertices are all on the same side of the plane formed by this set. But clearly, the amount of work to do this grows without bound as the number of vertices and/or the dimension of the polytope become large. 

Thus, what it means to ``understand" infinite families of polytopes, is to be able to analytically describe them in {\it both} the convex hull and facet descriptions. As we now show, the Symanzik polytopes can be ``understood" in this sense, and can be given an extremely simple facet presentation.

\paragraph*{\bf Inequalities.}
$\UU_G$ is the Newton polytope of a homogeneous degree $\LL_G$ polynomial and therefore is naturally defined on the hyperplane $a_G = \LL_G$, where $\vec a$ is a point in the Newton polytope and $a_\gamma = \sum_{e \in \gamma}a_e$.
On this hyperplane, $\UU_G$ is defined by
\begin{equation}
\label{udefn}
    \UU_G\!:  \quad a_G = \LL_G \quad\mathrm{and}\quad a_\gamma \geq \LL_\gamma \quad\text{for all } \gamma\text{'s}.
\end{equation}
Analogously, $\FF_G$ is cut out by
\begin{align}
\label{fdefn}
    \FF_G\!:\quad  a_G = \LL_G{+}1  \quad\text{and}\;\, &\begin{cases}
    a_\gamma \geq \LL_\gamma & \text{if}\quad\mathcal{F}_{G/\gamma} \neq 0,
    \\
    a_\gamma \geq \LL_\gamma{+}1 &  \text{if}\quad\mathcal{F}_{G/\gamma} = 0,
    \end{cases}\nn\\
    & \qquad\qquad\qquad\;\text{for all } \gamma\text{'s}.
\end{align}
Graphs $G/\gamma$ with $\F_{G/\gamma} = 0$ are called \emph{scaleless} and define inequalities crucial for the study of IR divergences. In contrast, for generic kinematics (say, all particles massive), the definition of $\FF_G$ only differs from that of $\UU_G$ by the choice of hyperplane and indeed is the Minkowski sum $\FF_G = \UU_G \oplus {\bf\Delta}_{\E-1}$, where ${\bf\Delta}_{\E-1}$ is the $(\E{-}1)$-dimensional simplex.

\paragraph*{\bf Boundary structure.}

Saturating the above inequalities defines faces of $\UU_G$ and $\FF_G$ with the following factorization properties:
\begin{equation}
\partial\UU_G \supset \UU_\gamma \times \UU_{G/\gamma} \quad\text{on}\quad a_\gamma = \LL_\gamma
\end{equation}
and
\begin{align}\label{eq:del-F}
    \partial\FF_G \supset
    \begin{cases}
    \UU_\gamma \times \FF_{G/\gamma} & \text{on}\quad a_\gamma = \LL_\gamma,\\
    \FF_\gamma \times \UU_{G/\gamma} & 
    \text{on}\quad a_\gamma = \LL_\gamma{+}1.
    \end{cases}
\end{align}
It follows that $\SS_G$ factorizes in the same way as $\FF_G$. Furthermore, we claim that \emph{facets} (codimension-$1$ faces) of $\SS_G$ are those faces of \eqref{eq:del-F} for which every instance of $\UU_{\gamma'}$ in the factorization above has $\gamma'$ 1VI (recall that a diagram is called 1VI if it cannot be disconnected by removal of a single vertex) and, in the $\F_{G/\gamma} = 0$ case, no subgraph $\tilde \gamma \subset \gamma$ already has $\F_{G/\tilde \gamma} = 0$ and $\L_{\tilde\gamma} = \L_\gamma$.

\paragraph*{\bf Compatibility of facets.} Feynman integrals motivate us to study the properties of $\SS_G$.  In particular, it is natural to ask if two facets labelled by $\gamma_1$ and $\gamma_2$ are compatible, i.e., they meet.  The following identity
\begin{align}
    a_{\gamma_1}+a_{\gamma_2} = a_{\gamma_1 \cup \gamma_2}+a_{\gamma_1 \cap \gamma_2}
\end{align}
furnishes a necessary condition for compatibility:
\begin{multline}
\label{compatibility}
     (1+c)(\L_{\gamma_1}+\L_{\gamma_2}) +(c)_{\gamma_1}+(c)_{\gamma_2} \geq\\
     (1+c)(\L_{\gamma_1\cup\gamma_2}+\L_{\gamma_1\cap\gamma_2}) +(c)_{\gamma_1\cup\gamma_2}+(c)_{\gamma_1\cap\gamma_2}.
\end{multline}
The $c$'s in the parentheses are only included when the $\FF_G$ facet is of $a_\gamma \geq \L_\gamma{+}1$ type.

Note that the above condition means Feynman polytopes are in general \emph{not} generalized permutohedra, in contrast with \cite{Brown:2015fyf,Schultka:2018nrs,Borinsky:2020rqs}.

\section{Tropicalization}

Recall that rays in the Trop space are dual to facets of the corresponding polytope.
Given the explicit form of the $\text{Trop}$ function \eqref{tropdef}, one can simply plug in the ray corresponding to a given inequality and get the associated value of $\text{Trop}$ along that ray.  However, equipped with the explicit \textit{constants} in the inequalities for $\UU_G$ and $\FF_G$, we already know the value of Trop along the facet rays: they are given by the constant in the inequality.

\paragraph*{\bf Degrees of divergence.} The two polytopes are defined by intersecting the same inequalities with different subspaces.  Because of projective invariance, we can uniformly shift the points on $\UU_G$ and $\FF_G$, which has no effect on the $\Trop$ function, with
\begin{align}
    \label{hyperplaneshifts}
    \vec a \to  \vec a- \tfrac{\LL_G}{\E_G}\vec 1_G, \hspace{7mm}  \vec a \to  \vec a-\tfrac{\LL_G+1}{\E_G}\vec 1_G,
\end{align}
respectively,
where $\vec 1_G$ is a vector with all entries equal to $1$. At the level of the integrand, this is equivalent to distributing the overall $\alpha_1\alpha_2\cdots \alpha_\E$ between the $\U_G$ and $\F_G$.  Note that no change of integration variables is implied here, but the inequalities in \eqref{udefn} and \eqref{fdefn} are shifted.
The resulting inequalities for $\UU_G$ are:
\begin{align}
    \label{shiftedineqsU}
    a_\gamma  \geq \frac{\LL_\gamma \E_G-\LL_G \E_\gamma}{\E_G}
\end{align}
and for $\FF_G$:
\begin{align}
\label{shiftedineqsF}
a_\gamma \geq
   \begin{cases}
   \frac{\LL_\gamma \E_G-(\LL_G+1)\E_\gamma}{\E_G}& \text{if}\quad\mathcal{F}_{G/\gamma} \neq 0,\\
    \frac{(\LL_\gamma+1)\E_G-(\LL_G+1)\E_\gamma}{\E_G}&\text{if}\quad\mathcal{F}_{G/\gamma} = 0,
   \end{cases}
\end{align}
for the two types of $\FF_G$ facets.  The value of Trop along a ray is given by substituting the saturation of that ray's inequality in the maxima. Since our conventions identify the value of Trop with the constant in the outward-pointing normal inequality, we substitute the constants in \eqref{shiftedineqsU} and \eqref{shiftedineqsF} with a minus sign, see App.~\ref{app:general}. This allows us to compute the value of Trop along any extremal ray.  The two types of faces are distinguished by their inequality in $\FF_G$ and give two cases:
\begin{align}\label{eq:Trop-gamma}
    \text{Trop} = \begin{cases}
    -d_\gamma & \text{if}\quad \mathcal{F}_{G/\gamma} \neq 0, \\
    d_{G/\gamma} & \text{if}\quad \mathcal{F}_{G/\gamma} = 0.
    \end{cases}
\end{align}
The first case is familiar and associated with superficially UV divergent subgraphs when $d_\gamma \leq 0$.  We will see that the second case is associated with IR divergences when $d_{G/\gamma} \geq 0$ and $G/\gamma$ is scaleless.

\section{Generalized Weinberg's theorem}
The above geometric picture makes Weinberg's theorem \cite{PhysRev.118.838} obvious: if all subdiagrams $\gamma \subset G$ have $d_{\gamma}>0$ (with appropriately defined $d_\gamma$ for non-scalar diagrams), the diagram $G$ is UV convergent.
With \eqref{eq:Trop-gamma}, we are now equipped to state its IR counterpart: if all subdiagrams $\gamma \subset G$ for which $G/\gamma$ is scaleless have $d_{G/\gamma} < 0$, the diagram $G$ is IR finite. Together, the two theorems give conditions for any diagram to be UV/IR finite, cf. \cite{Lowenstein:1975rg}.

\section{\label{sec:UV} Ultraviolet}
\begin{figure}[!t]
	\centering
	\includegraphics[]{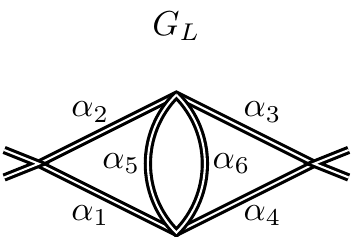}
	\quad
    \raisebox{-.15\height}{\includegraphics[]{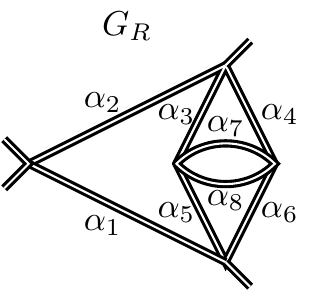}}
	\caption{\label{eyeofsauron}Two UV-divergent Feynman diagrams $G_L$ and $G_R$ considered in Sec.~\ref{sec:UV}.
	}
\end{figure}

We now provide some illustrative examples.  As a UV example, we consider the diagram $G_L$ in Fig.~\ref{eyeofsauron} (left). It contributes a leading logarithmic divergence in $\D = 4-2\varepsilon$ and therefore has a two-dimensional divergent Trop = 0 space (recall that one divergence comes from the overall $\Gamma(d_{G_L}) = \tfrac{1}{3\varepsilon} + \dots$).  In order to calculate the leading divergence, we first need to identify the divergent rays spanning the Trop = 0 space.  These correspond to the superficially divergent subgraphs, of which there are three: $\gamma_{1256}$, $\gamma_{3456}$, $\gamma_{56}$ where the subscripts denote the edges of the subgraph.  Next, we need to know the compatibility of these divergent rays in $\SS_{G_L}$.  This is given by the compatibility criterion (\ref{compatibility}) and demonstrates that $\gamma_{56}$ is compatible with each of the other two subgraphs, but $\gamma_{1256}$ and $\gamma_{3456}$ are not compatible with each other.  This means that we have two two-dimensional cones and they share a ray.  We must compute the contribution to the leading divergence from each two-dimensional cone separately and add them.

We may now leverage eq.~(\ref{tropvolume}) to obtain
\begin{align}
    \label{eyedivergence}
    \I_{G_L} = \Gamma(d_{G_L}) \frac{1}{d_{\gamma_{56}}}\left(\frac{1}{d_{\gamma_{1256}}}+\frac{1}{d_{\gamma_{3456}}}\right)+\ldots,
\end{align}
where the finite integral is accounted for as it is merely a product of bubble integrals which equal one. Since $d_{\gamma_{1256}} = d_{\gamma_{3456}} = 2\varepsilon$ and $d_{\gamma_{56}} = \varepsilon$, the leading UV divergence contributes $\I_{G_L} = \frac{1}{3\varepsilon^3} + \ldots$.  The structure of the rays in this example captures the familiar RG intuition of sequentially shrinking loops.  One simplicial cone is generated by rays $r_{1256}$ and $r_{56}$ and captures the shrinking of the two corresponding subgraphs, one nested in the other.  Any ray in this cone corresponds to a direction of scaling integration variables where the bubble $\gamma_{56}$ shrinks at least as fast as the collective $\gamma_{1256}$, capturing the physics of shrinking successive UV subgraphs to effective vertices.

\begin{figure}[!t]
    \centering
    \includegraphics[]{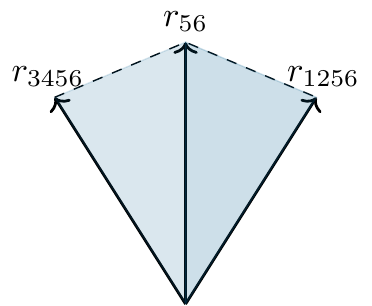}
	\includegraphics[]{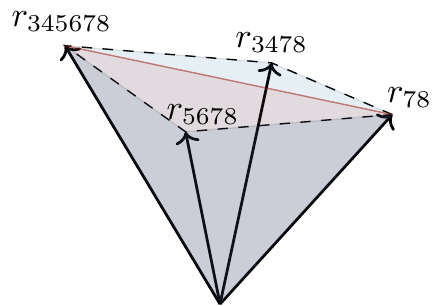}
    \caption{Divergent rays of the diagrams from Fig.~\ref{eyeofsauron}. Left: Two two-dimensional cones which share a single ray $r_{56}$, reflecting the fact that there is an overlapping divergence. Right: Triangulation of the three-dimensional cone into two simplicial cones separated by the red plane.  In this triangulation each simplex has the interpretation of nested shrinking subgraphs with ordering implied by the rays.}
    \label{eyeinbubblerays}
\end{figure}

\paragraph*{\bf Leading divergences and the RG.}
In general, the maximal cones Trop = 0 cones need not be simplicial. A given cone's contribution then need not be interpreted in the familiar nested shrinking picture consistent with our familiar RG intuitions.  Nonetheless, there is a triangulation of the cone that does this.  Consider the diagram $G_R$ from Fig.~\ref{eyeofsauron} (right), whose rays are depicted in Fig.~\ref{eyeinbubblerays} (right).  The triangulation divides the cone into two simplices, with rays that again imply an ordering of scaling rate.  That is, any ray in the simplex generated by $r_{345678}, r_{3478}, r_{78}$ has components $\gamma_{78}$ shrink the fastest, followed by $\gamma_{34}$ and then $\gamma_{56}$.  The nested shrinking is therefore one natural way of organizing the calculation of the leading the divergence, though not uniquely so.  Using \eqref{tropvolume}, we can verify that the two triangulations produce the same leading contribution:
\begin{align}
    &\frac{1}{d_{\gamma_{345678}}d_{\gamma_{78}}}\left(\frac{1}{d_{\gamma_{3478}}}+\frac{1}{d_{\gamma_{5678}}}\right)\nn\\ =&\,\frac{1}{d_{\gamma_{3478}}d_{\gamma_{5678}}}\left(\frac{1}{d_{\gamma_{78}}}+\frac{1}{d_{\gamma_{345678}}}\right) = \frac{1}{3\varepsilon^3},
\end{align}
and hence the leading contribution in this case is $\mathcal{I}_{G_R} = \Gamma(d_{G_R}) \frac{1}{3\varepsilon^3} + \ldots = \frac{1}{12\varepsilon^4}+\dots$.
Similarly, the phenomenon of overlapping divergences for the diagram $G_L$ is illustrated in Fig.~\ref{eyeinbubblerays} (left), where the ray $r_{56}$ is shared between two two-dimensional cones.

The fact that leading UV divergences are organized in a ``Russian doll" structure of shrinking nested subgraphs, reflects the Wilsonian way of organizing the generation of logarithmic divergences scale-by-scale. This is a simple but important first step in understanding the origin of the renormalization group. Of course the full structure of the RG goes beyond this basic fact about leading divergences of individual diagrams, and organizes the UV divergences into leading, subleading etc. logarithms captured by running coupling constants. We will discuss a geometric perspective on this understanding in \cite{AHHM}.

\section{\label{sec:IR} Infrared}

\begin{figure}[!t]
    \centering
    \includegraphics[scale=.7]{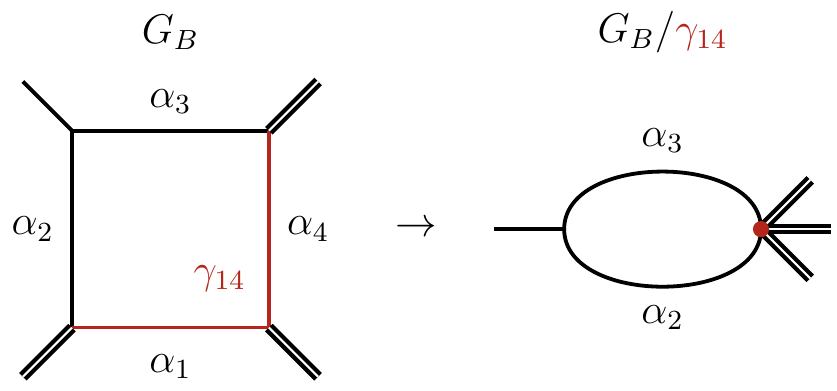}
    \caption{Left: Collinear divergence illustrated for the box diagram $G_B$ with one massless corner (single line) and the remaining ones massive (doubled line). A subgraph $\gamma_{14}$ corresponding to the graph's unique divergent ray in $\D = 4-2\varepsilon$.  Right: The resulting $G_B/\gamma_{14}$ is scaleless.}
    \label{IRcorner}
\end{figure}

Combinatorics of leading divergences is much richer in the IR.
The rays for which $\text{Trop} = -d_{G/\gamma}$ correspond to scaleless $G/\gamma$'s.  We are particularly interested in $G/\gamma$'s for which  $d_{G/\gamma} = 0$, as they correspond to logarithmic IR divergences.
Such $G/\gamma$'s can then be blown up into a graph $G$ by inserting a $\gamma$ into it. By studying the possible structures of $G/\gamma$'s, we can understand the ways in which IR divergences can appear.

All scaleless graphs must have massless internal edges.  There are then two ways to make scaleless graphs: momentum conservation or masslessness.  Either the only spanning $2$-trees isolate a single massless external momentum or they isolate all external momenta and vanish by momentum conservation.  In order to have a divergence, the sign of Trop tells us that  we want $d_{G/\gamma}$ to be as convergent as possible; the more superficially convergent $d_{G/\gamma}$, the more divergent the contribution to the Feynman integral $\I_G$.  But we also need $G/\gamma$ to be scaleless. There is a basic tension between being scaleless and being superficially convergent.  We can make a graph arbitrarily superficially convergent by adding external lines, i.e., bivalent vertices in the amputated graph, but we cannot have more than two such vertices with external kinematics and be scaleless. 

\begin{figure}[!t]
    \centering
    \includegraphics[]{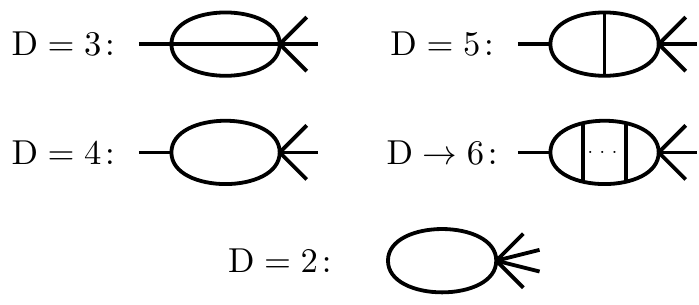}
    \caption{Leading $G/\gamma$ graphs which are scaleless and marginal in the relevant dimensions.  We write $\D\to 6$ as marginality is only attained in the limit of an infinite number of rungs.}
    \label{IRdimensions}
\end{figure}

\paragraph*{\bf In which dimensions do IR divergences exist?}
This discussion allows us to bound the dimensions in which IR divergences appear.  In particular, we can ask: what is the highest dimension in which a marginal IR divergence may exist?  For $n > 3$, we will maximize $\E-\LL \D/2$ when we have a bubble filled with trivalent vertices.  In this case, with $2v$ trivalent vertices, we have 
\begin{align}
    d_{G/\gamma} = \frac{(4-\D) + v(6-\D)}{2}.
\end{align}
We see that as $v\to \infty$ then $\D \to 6$ becomes marginal, therefore one will not find IR divergences in six or higher dimensions.
Since adding numerators can only soften IR divergences, these bounds apply to a general theory.
Fig.~\ref{IRdimensions} catalogues the leading examples of marginal scaleless graphs in the relevant dimensions.

It is worth noting that the existence of marginal IR divergences in a given dimension implies the existence of power IR divergences in lower dimensions.
In particular, scalar theories such as $\phi^3$ exhibit power IR divergences in $\D=4$ \cite{AHHM}, see Fig.~\ref{fig:scalings}c, giving an additional raison d'être for numerators in gauge theories: softening IR divergences to logarithmic.

\paragraph*{\bf Soft/collinear divergences.}

The standard collinear divergence comes from a ray $r_{14}$ corresponding to shrinking $\gamma_{14}$ (or expanding $\gamma_{23}$), as illustrated in Fig.~\ref{fig:scalings}a. It is singular because $G_B/\gamma_{14}$ is scaleless, see Fig.~\ref{IRcorner}. 
The cone is therefore a single ray $r_{14}$, which according to \eqref{tropvolume} gives, in the conventions of Fig.~\ref{IRcorner}
\begin{align}\!
\vcenter{\hbox{\includegraphics[width=7mm]{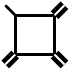}}}\!\!&=\!  -\frac{\Gamma(d_{G_B}\!)}{d_{G_{\!B}\!/\gamma_{14}}}\!\! \int\!\!\! \frac{\d^4 \alpha/\text{GL(1)}^2}{(s\alpha_1\alpha_3{+}t\alpha_2\alpha_4{+}p_1^2\alpha_1\alpha_2{+}p_3^2\alpha_3\alpha_4)^2} \!+\!\dots\nn\\
&= -\frac{1}{\varepsilon}  \int_{0}^{\infty}\!\! \frac{\d \alpha_1 \d \alpha_2 }{(s \alpha_1+t\alpha_2+p_1^2\alpha_1\alpha_2+p_3^2)^2} + \dots,\nn\\
&= -\frac{1}{\varepsilon}  \frac{\log\left(\frac{st}{p_1^2 p_3^2}  \right)}{s t-p_1^2p_3^2 } +\dots,
\label{eq:one-corner-box}
\end{align}
where we used the two $\mathrm{GL}(1)$'s to fix $\alpha_3 = \alpha_4 = 1$ and $d_{G_B} = 2{+}\varepsilon$ with $d_{G_B/\gamma_{14}} = \varepsilon$,
see App.~\ref{app:further-examples} for more details.

The soft/collinear divergence is illustrated in Fig.~\ref{fig:scalings}b (see also \cite{YelleshpurSrikant:2019khx}) and comes from a two-dimensional cone generated by two IR rays, say $r_{14}$ and $r_{34}$, sharing an edge. Its leading divergence is
\begin{align}
\vcenter{\hbox{\includegraphics[width=7mm]{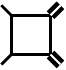}}} &= \frac{\Gamma(d_{G_B})}{d_{G_B\!/\gamma_{14}}\, d_{G_B\!/\gamma_{34}}} \int  \frac{\d^4 \alpha/\text{GL(1)}^3}{(s\alpha_1\alpha_3+t\alpha_2\alpha_4)^2} + \dots \nn\\
&= \frac{1}{\varepsilon^2 st} +\dots,
\end{align}
where we used the enhanced $\mathrm{GL}(1)^3$ to fix $\alpha_2=\alpha_3=\alpha_4=1$. Generalizations to nested IR divergences can be quite intricate, see, e.g., Fig.~\ref{fig:scalings}d.

\paragraph*{\bf Fishnet diagrams.} 

\begin{figure}[!t]
    \centering
    \scalebox{0.5}{
    \includegraphics[]{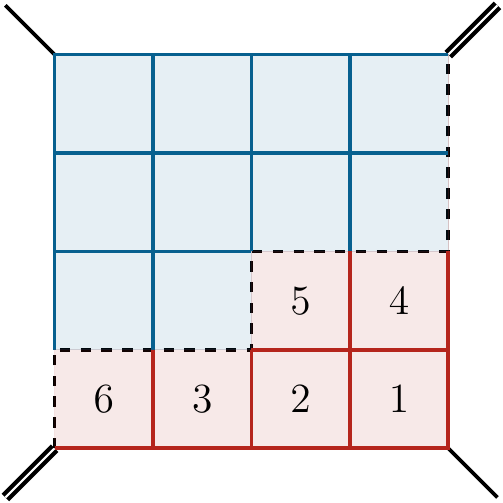}}
    \hspace{1.25cm}\scalebox{1.15}{
   \includegraphics[]{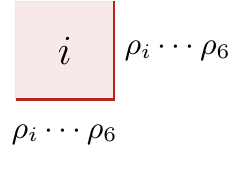}}
    \caption{Left: Half of a top-dimensional contribution to a four-point fishnet diagram with opposite massless corners.  The IR subgraphs (red and blue plus respective sub-tableaux) are complementary to those appearing in the inequalities, i.e., these are the edges becoming long/soft.  We additionally label for the red portion one of the standard tableaux corresponding to a cell in the triangulation. Right: Scaling parameters $\rho_i \to \infty$ of the lower-right edges for a square labeled by integer $i$ in the standard tableaux.}
    \label{fishnetexample}
\end{figure}

For scalar integrals in $\D=4$, it is natural to study the generalization to \emph{fishnet} diagrams, i.e., planar arrays of $M \times N$ squares \cite{Gurdogan:2015csr,Chicherin:2017frs,Loebbert:2020tje}, see Fig.~\ref{fishnetexample} for $M=N=4$ with two massless corners.
As discussed in Sec.~\ref{sec:general},
calculation of the divergent contribution from a graph is performed by identifying the rays generating divergent cones and applying \eqref{tropvolume}.  In the case of IR-divergent fishnets, the combinatorics of these rays is rich, but admits a simple description.  It suffices to study the case with two diagonally-opposed massless corners, as the combinatorics of IR rays associated with neighboring corners is trivial, i.e., the geometry is a product.  This means that for any assignment of the external masses, we can treat the diagonals independently and multiply the result.  A single leading-dimensional cone of the Trop = 0 space is in correspondence with a staircase walk (dashed lines) separating the fishnet into two Young tableau-shaped sets of edges (blue and red) as in Fig.~\ref{fishnetexample}. 
Then the compatible rays $\gamma$ are given by the complements of 
all the Young \emph{subdiagrams} based at the top-left (blue) or bottom-right corner (red), including the dashed bordering edges.  For example, the blue shading with the dashed lines is the $\gamma$ corresponding to the tableaux given by the red shading.

Since such cones are in general not simplicial, in order to leverage \eqref{tropvolume} one needs a triangulation.  For concreteness, consider the red subdiagram in Fig.~\ref{fishnetexample}.  One triangulation is given by cells corresponding to the \emph{standard} tableaux, see Fig.~\ref{fishnetexample}. Scaling of the bottom and right edges of every red box labeled with integer $i$ is given by $\rho_i \rho_{i+1} \cdots \rho_6$, where each $\rho_i$ parameterizes scaling along a ray generating the simplicial cell.  Each cell clearly contributes the volume $\frac{1}{6!}$, so the non-trivial combinatorics is in the counting of the standard tableaux, given by the hook-length formula.  We claim that summing over the volume contributions from all the cells yields
\begin{equation}
    \label{catalan}
    \frac{2^{MN}}{(MN)!} C_{M,N} = \!\!\sum\limits_{\lambda_{i} > \lambda_{i+1}}\frac{f^{(\lambda_1, \dots, \lambda_M)}f^{(\bar \lambda_M, \dots, \bar \lambda_1)}}{(\lambda_1{+}\dots{+} \lambda_M)!(\bar\lambda_1{+}\dots{+} \bar\lambda_M)!}  ,
\end{equation}
where $\bar \lambda_i = N-\lambda_i$ and $f^{(\lambda_1, \dots, \lambda_M)}$ counts the number of the standard Tableaux with shape $\lambda = (\lambda_1, \dots, \lambda_M)$, and $C_{M, N}$ is the multi-dimensional Catalan number \cite{MultiCatalan} with standard definition:
\begin{equation}
    C_{M,N} = (M N)!\frac{\text{sf}(M-1)\text{sf}(N-1)}{\text{sf}(M+N-1)}
\end{equation}
with the superfactorial $\text{sf}(N) = 0!1!2!\cdots N!$.

For the fishnet with one massless corner, there is no sum over partitions and we merely have to count the number of standard tableaux for the $M \times N$ rectangle, which is given by $C_{M, N}$:
\begin{equation}
\label{3massfish}
\raisebox{-2em}{\includegraphics[]{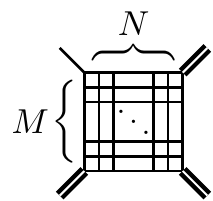}}
\;=\; \frac{F_{M,N}(s, t, p_i^2)}{(MN)!} \frac{C_{M,N}}{\varepsilon^{MN}} + \ldots,
\end{equation}
where $F_{M,N}(s, t, p_i^2)$ is a non-trivial function of kinematics computed by the $\text{GL}(1)^{MN}$-modded integral in the way described above, see \eqref{eq:one-corner-box} for a special case. And in the case of two diagonally opposed massless edges, we employ \eqref{catalan} yielding:
\begin{equation}
\label{2massfish}
\raisebox{-2em}{\includegraphics[]{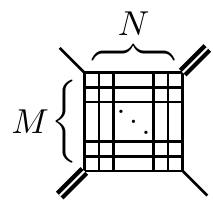}}
\;=\; \frac{2^{MN}}{(MN)!} \frac{C_{M, N}}{\varepsilon^{MN} s^N t^M} + \ldots,
\end{equation}
The leading contribution to the fully massless fishnet is obtained by squaring the kinematic-independent part of \eqref{2massfish} and keeping the same kinematic dependence.\footnote{Similarly, the contribution with two neighboring massless corners is obtained by squaring the kinematic-independent part of \eqref{3massfish} and doing the appropriate modded integral giving $1/(s^N t^M)$. The contribution with the three massless corners is obtained by multiplying kinematic-independent parts of \eqref{3massfish} and \eqref{2massfish} and again obtaining $1/(s^N t^M)$ from the modded integral.} See \cite{AHHM} for details and \cite{Basso:2021omx,Olivucci:2021pss} for alternative perspectives.

\section{Outlook}

In this work we introduced polytopes capturing the structure of UV/IR divergences of Feynman integrals. Recently, Symanzik polytopes also played a role in the study of kinematic singularities, known as anomalous thresholds \cite{Mizera:2021icv}, and it would be interesting to relate the two approaches. Other future direction include a systematic approach to extracting subleading divergences, inclusion of kinematic numerators, as well as summing over multiple Feynman diagrams under the same integral, further pursued in \cite{AHHM}.

\begin{acknowledgments}
We thank Francis Brown for introducing us to Symanzik polytopes and for sharing his unpublished notes \cite{BrownIHES}. We also thank Nick Early, Hofie Hannesdottir, Song He, Thomas Lam, Erik Panzer,  Giulio Salvatori, and Simon Telen for stimulating discussions. This work is supported by the grant DE-SC0009988 from the U.S. Department of Energy.
S.M. gratefully acknowledges the funding provided by Frank and Peggy Taplin. 
\end{acknowledgments}

\appendix
   
\section{\label{app:general}Leading divergence of general integrals}

Let us consider a family of integrals which can be written as
\begin{equation}\label{eq:I-appendix}
\I = \int_{\mathbb{R}_+^d} \d^{d} X \prod_{i=1}^{d} X_i^{a_i-1} \prod\limits_{j} P_j(X)^{-c_j},
\end{equation}
where $P_j = \sum_{k} c_{k}^{(j)} \prod_{l=1}^{d} X_i^{v_{i}^{(j)}}$ are $m$ arbitrary polynomials in $d$ integration variables $X_i$. Feynman integrals \eqref{eq:I} are a special case of \eqref{eq:I-appendix}.
We want to understand singularities of such integrals in the parameters $a_i$ and $c_j$, 
leveraging the tropical geometry associated to their integrands via the Newton polytope map.

The change of variables $X_i = e^{x_i}$ naturally motivates tropicalization for the study of the asymptotic regions of the integration domain, $\mathbb{R}^d$.  In particular, when the magnitude of the $x_i$'s scales to infinity, we are left to study rays in $\mathbb{R}^d$, where the direction characterizes the relative scaling rates of the $x_i$'s. The \textit{leading} asymptotic behavior of the integrand along a given ray is $e^{\Trop}$ with
\begin{equation}
\Trop(x) = a\cdot x-\sum\limits_j c_j \max ( x\cdot v^{(j)} ),
\end{equation}
where the dot denotes the inner product on $\mathbb{R}^d$ and the max condition is the maximum over all monomials $v$ in a given polynomial $P_j$. The $\Trop$ function captures the leading asymptotic behavior of the integrand, and we can compute the leading contribution to the integral from this.

Along a given ray $x$, there is at least one maximal monomial $x \cdot v$ such that $x \cdot v \geq x \cdot v'$ for any other $v'$.  The $d$ dimensional set of rays $x$ obeying this inequality comprise a domain of linearity of $\Trop$, which is a cone we call $C_v$.  The whole integration region $\mathbb{R}^d$ is tiled by such cones, each of which is generated by a set of extremal rays $w_\alpha^{(v)}$ such that inside the cone $C_v$ we have
\begin{equation}\label{eq:x-lambda}
    x = \sum_{\alpha} \lambda_\alpha w_\alpha^{(v)} \quad\text{with}\quad \lambda_\alpha > 0.
\end{equation}
Because the $\Trop$ function is linear inside of a cone, knowledge of the extremal rays $w_\alpha^{(v)}$, together with the cones $C_v$ which they bound and the values $\Trop(w_\alpha^{(v)})$, determines the values of $\Trop(x)$ everywhere. 

In order to compute the leading divergence, without loss of generality we can focus on the contribution from one simplicial cone (i.e., when the number of $w_\alpha^{(v)}$'s is exactly $d$), as the full answer is obtained by summing over triangulations of such cones. When the contribution is top-dimensional, we utilize the change of variables from $x$ to $\lambda$, as defined in \eqref{eq:x-lambda}.
The contribution from such a cone $C_v$ is then given by
\begin{align}\label{eq:I-simplicial}
     \I_{C_v} &= |w_1^{(v)}\cdots w_d^{(v)}|\int_{\mathbb{R}^d_+} \d^d\lambda \; e^{\sum_{\alpha} \lambda_\alpha \Trop(w_\alpha^{(v)})} \nn\\
    &= (-1)^d \frac{|w_1^{(v)}\cdots w_d^{(v)}|}{\Trop(w_1^{(v)})\cdots \Trop(w_d^{(v)})},
\end{align}
where the numerator features the Jacobian determinant.
In other words, the leading behavior is given by the volume of the cone bounded by $\Trop(x) \geq -1$ inside $C_v$. In practice, what is useful for us is choosing vectors $w_i^{(v)}$ such that the Jacobian determinant is one in which case the leading behavior is just the inverse product of $\Trop$ along those rays.
 
Finally, in order to apply this formalism to the study of Feynman integrals, we need to understand contributions from lower-dimensional cones, where there remains integration off the cone to be done.  In this case, we can parameterize a ray $x$ in an $m$-dimensional cone $C_F$ as 
 \begin{equation}
     x = \sum_{\alpha}\lambda_\alpha w_\alpha^{(F)} + \sum_{\beta} \bar \lambda_\beta \bar w_\beta^{(F)} \quad\text{with}\quad \lambda_\alpha > 0,
 \end{equation}
 where the $\bar w_\beta^{(F)}$'s are orthogonal to the span of the $w_\alpha^{(F)}$'s and the $\bar \lambda_\beta$ can have any real value because we are interested in the contribution to the integral dominated by large $\lambda_\alpha$.  The integration regions decouple between the $\lambda_\alpha$ and $\bar \lambda_\beta$, and because we only keep monomials that are leading on this cone, the integrand factors as well. Once again considering the simplicial case, we have
 \begin{align}
    \I_{C_{F}} &= \left|w_1^{(F)}\cdots w_m^{(F)} \bar w_1^{(F)} \cdots \bar w_{d-m}^{(F)} \right|\\ &\quad\times\int_{\mathbb{R}^m_+} \d^m\lambda\; e^{\sum_{\alpha} \lambda_\alpha \Trop(w_\alpha^{(F)})} \nn\\
    &\quad\times \int_{\mathbb{R}^{d-m}}\!\!\!\!\! \d^{d-m} \bar \lambda \; e^{\sum_{\beta} \bar \lambda_\beta a\cdot \bar w_\beta^{(F)}} \prod_j P_j(e^{\sum_{\beta} \bar\lambda_\beta \bar w_\beta^{(F)}})^{-c_j}.\nn
\end{align}
As before, the first two lines are simply the volume 
\begin{equation}\label{eq:volC}
\mathrm{vol}(C_{F}) = (-1)^m \frac{\left|w_1^{(F)}\cdots w_m^{(F)} \bar w_1^{(F)} \cdots \bar w_{d-m}^{(F)} \right|}{\Trop(w_1^{(F)})\cdots \Trop(w_m^{(F)})}.
\end{equation}
The final line is a finite integral $\J_{F}$ over $\mathbb{R}^{d-m}$ that remains off of the $m$-dimensional cone.  This finite integral can also be understood as coming from gauge-fixing the enhanced $\mathrm{GL}(1)^{m}$ symmetry found on the integrand keeping only the leading monomials. We use the shorthand notation
\begin{equation}
\J_{F} = \int \frac{\d^d  \alpha}{\mathrm{GL}(1)^{m}}\, \prod_{i=1}^{d} X_i^{a_i-1} \prod\limits_{j} P_j(X)^{-c_j} \bigg|_{C_F^\perp},
\end{equation}
where $C_F^\perp$ denotes the orthogonal directions to the cone $C_F$. Note that $\mathrm{vol}(C_{F})$ and $\J_F$ depend on the choice of splitting the space into $C_F$ and $C_F^\perp$ (through the Jacobian), but their product is invariant. 

In summary, rescaling all the exponents $a_i {\to} \varepsilon a_i$, $c_j {\to} \varepsilon c_j$, the leading behavior of \eqref{eq:I-appendix} as $\varepsilon \to 0$ is given by
\begin{equation}
\I =  \sum_{F} \mathrm{vol}(C_{F}) \J_F + \ldots,
\end{equation}
where the sum goes over all $\Trop=0$ cones $C_F$ of the maximal dimension $m$, such that each $\mathrm{vol}(C_F) \propto 1/\varepsilon^m$.

We have reduced the problem of calculating the leading divergence to determining the values of the \text{Trop} along the extremal rays of the divergent cones and performing an additional finite integral.  It remains to explain the connection to the polytopes studied in this letter.  In the case of one polynomial, the monomial associated with cone $C_F$ is a vertex of the Newton polytope
\begin{equation}
c_j\, \Newt(P_j).
\end{equation}
This is because it obeys the inequality $x\cdot v \geq x\cdot v'$ and, indeed, along an extremal ray bounding the cone, this vertex must saturate a facet inequality, say $w_\alpha^{(F)} \cdot v = c_\alpha$. Comparing to the $\Trop$ function we have $\Trop(w_\alpha^{(F)}) = c_\alpha$. While the vertex description of these polytopes is given by the Newton polytope definition, finding its facet description would give us the values of the $\Trop$ function on all extremal rays and therefore allow us to calculate the leading behavior of the integral.

This extends to multiple polynomials as a Minkowski sum if all $c_j > 0$:
\begin{equation}
\SS = - \vec{a} \oplus \bigoplus_j c_j\, \Newt(P_j),
\end{equation}
which is oriented so that rays are \emph{outward} pointing normals.
When $c_j$'s have generic signs, our analysis is also unaffected.  Despite the lack of a good notion of a Minkowski difference, our analysis only relies globally on the combinatorial structure, which is independent of these signs, and then also on the geometry of individual cones, not on full convexity of the $\SS$.
The convexity of cones and well-definedness of their volumes is respected for generic signs of the $c_j$'s.  In other words, the quantities we compute can be calculated \textit{as if} the $c_j$ were positive and can be applied to the generic $c_j$'s relevant for the tropical analysis with no trouble.

\begin{figure}[!t]
    \centering
    \includegraphics[]{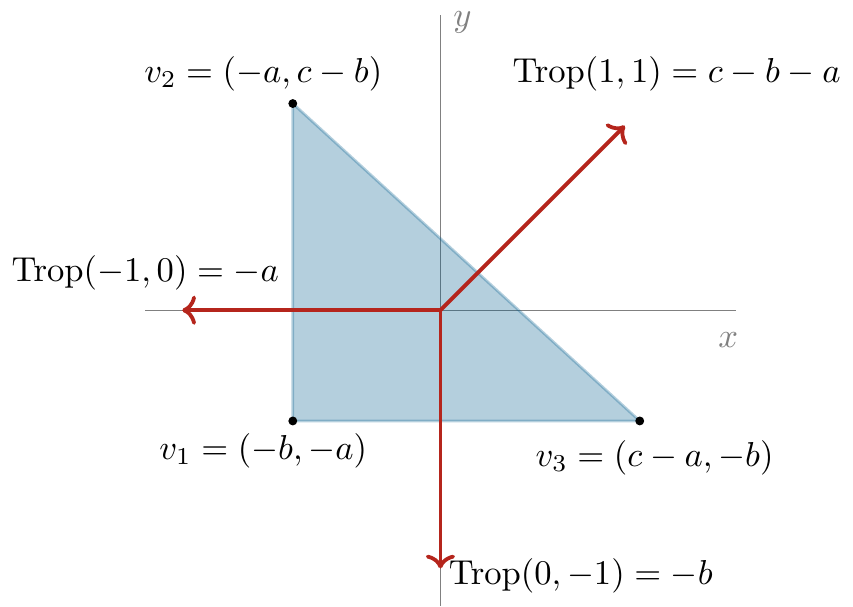}
    \caption{\label{fig:toy}The $\SS$ polytope (blue) and tropical fan (red) associated to the integral \eqref{eq:I-toy}. Each of the three rays has the corresponding value of the Trop function for the given vector along that ray.}
\end{figure}

\section{\label{app:further-examples} Further examples}

\noindent\par{\bf Toy integral.}
Let us begin with a toy example 
\begin{align}\label{eq:I-toy}
    \I_{\mathrm{toy}} &= \int_{\mathbb{R}_{+}^2} \frac{\d X \d Y}{X Y}X^a Y^b (1+X+Y)^{-c} \nn\\
    &= \frac{\Gamma(a)\Gamma(b)\Gamma(c-a-b)}{\Gamma(c)}.
\end{align}
The Trop function in this case is
\begin{equation}
    \Trop(x, y) = a x + by - c\, \text{max}(0, x, y).
\end{equation}
We see from the tropical fan that there are three extremal rays, see Fig.~\ref{fig:toy}. Consider taking the limit $a\to 0$ and $b \to 0$ with $c$ finite and positive. This corresponds to making two rays
\begin{equation}\label{eq:w1w2}
w_1^{(v_1)} = \begin{pmatrix}
    -1 \\ 0
    \end{pmatrix},\qquad
   w_2^{(v_1)} = \begin{pmatrix}
    0 \\ -1
    \end{pmatrix}
\end{equation}
divergent, thus generating a two-dimensional cone in the lower-left quadrant. According to the expression \eqref{eq:I-simplicial}, the leading contribution is given by the volume of the cone $C_{v_1}$ dual to the vertex $v_1$:
\begin{equation}
\I_{\mathrm{toy}} = \underbrace{\frac{\left|w_1^{(v_1)} w_2^{(v_1)} \right|}{\Trop(w_1^{(v_1)})\Trop(w_2^{(v_1)})}}_{\mathrm{vol}(C_{v_1})} + \ldots = \frac{1}{ab} + \ldots,
\end{equation}
which one can easily confirm by expanding the explicit answer \eqref{eq:I-toy}.

We can also consider only the limit $c-a-b \to 0$, in which we get a leading contribution from one ray $w_1^{(F)}$.  In this case, a finite integral remains. We pick the rays
\begin{equation}
    w_1^{(F)} = \begin{pmatrix}
    1 \\ 1
    \end{pmatrix},\qquad \bar w_1^{(F)} = \begin{pmatrix}
    \alpha \\ \beta
    \end{pmatrix}
\end{equation}
and leave the ray $\bar w_1^{(F)}$ off of the divergent cone generic in order to emphasize the result is independent of this choice. It is dual to the facet $F$ between the vertices $v_2$ and $v_3$. We have the leading contribution 
\begin{align}
   \I_{\mathrm{toy}} &= \underbrace{\frac{|\alpha-\beta |}{c-b-a}}_{\mathrm{vol}(C_{F})} \underbrace{\int_{\mathbb{R}} \d\bar\lambda_1\, e^{ (a\alpha+b \beta)\bar\lambda_1}\left(e^{ \alpha \bar\lambda_1}+e^{  \beta \bar\lambda_1} \right)^{-c}}_{\J_F} + \ldots
   \end{align}
Evaluating the integral cancels the Jacobian $|\alpha-\beta|$ and we find
 \begin{align}
   \I_{\mathrm{toy}} &= \frac{1}{c-b-a} \frac{\Gamma(a)\Gamma(b)}{\Gamma(c)} + \ldots,
\end{align}
again recovering the correct leading behavior.

\par{\bf One-mass triangle.} Note that the above integral is a Feynman integral in disguise. For instance, a massless triangle diagram with one massive corner is computed by
\begin{equation}
\I_{\mathrm{tri}} = \Gamma(3-\tfrac{\D}{2})\int \frac{\d^3 \alpha}{\mathrm{GL}(1)} \frac{(\alpha_1 + \alpha_2 + \alpha_3)^{3-\D}}{(p_1^2 \alpha_1\alpha_2 )^{3-\D/2}}.
\end{equation}
Using the $\mathrm{GL}(1)$ redundancy to fix $\alpha_3 = 1$, we obtain the integral \eqref{eq:I-toy} with $(X,Y)=(\alpha_1,\alpha_2)$ and
\begin{equation}
(a,b,c) = (\tfrac{\D-4}{2}, \tfrac{\D-4}{2}, \D{-}3),
\end{equation}
up to a prefactor. The two-dimensional cone generated by \eqref{eq:w1w2} in the $a,b\to 0$ limit now happens when $\D \to 4$. This is in fact the soft/collinear divergence at the two massless corners, associated to the limit in which the Schwinger parameters expand according to
\begin{equation}
(\alpha_1 : \alpha_2 : \alpha_3) = (\rho : \sigma : \rho \sigma) = (\tfrac{1}{\sigma} : \tfrac{1}{\rho} : 1)
\end{equation}
with $\rho,\sigma \to \infty$ (due to the overall scale redundancy, this IR singularity happens when $\alpha_1,\alpha_2 \to 0$ in the above $\mathrm{GL}(1)$ fixing). In terms of $\D=4{-}2\varepsilon$, the leading IR behavior is
\begin{equation}
\I_{\mathrm{tri}} = \frac{1}{\varepsilon^2 p_1^2} + \ldots.
\end{equation}
In the polytope language, the divergence came from the vertex (codimension-$2$ face) $v_1$ of $\SS$ in Fig.~\ref{fig:toy}.

\par
\begin{figure}[!t]
    \includegraphics[width=0.49\columnwidth]{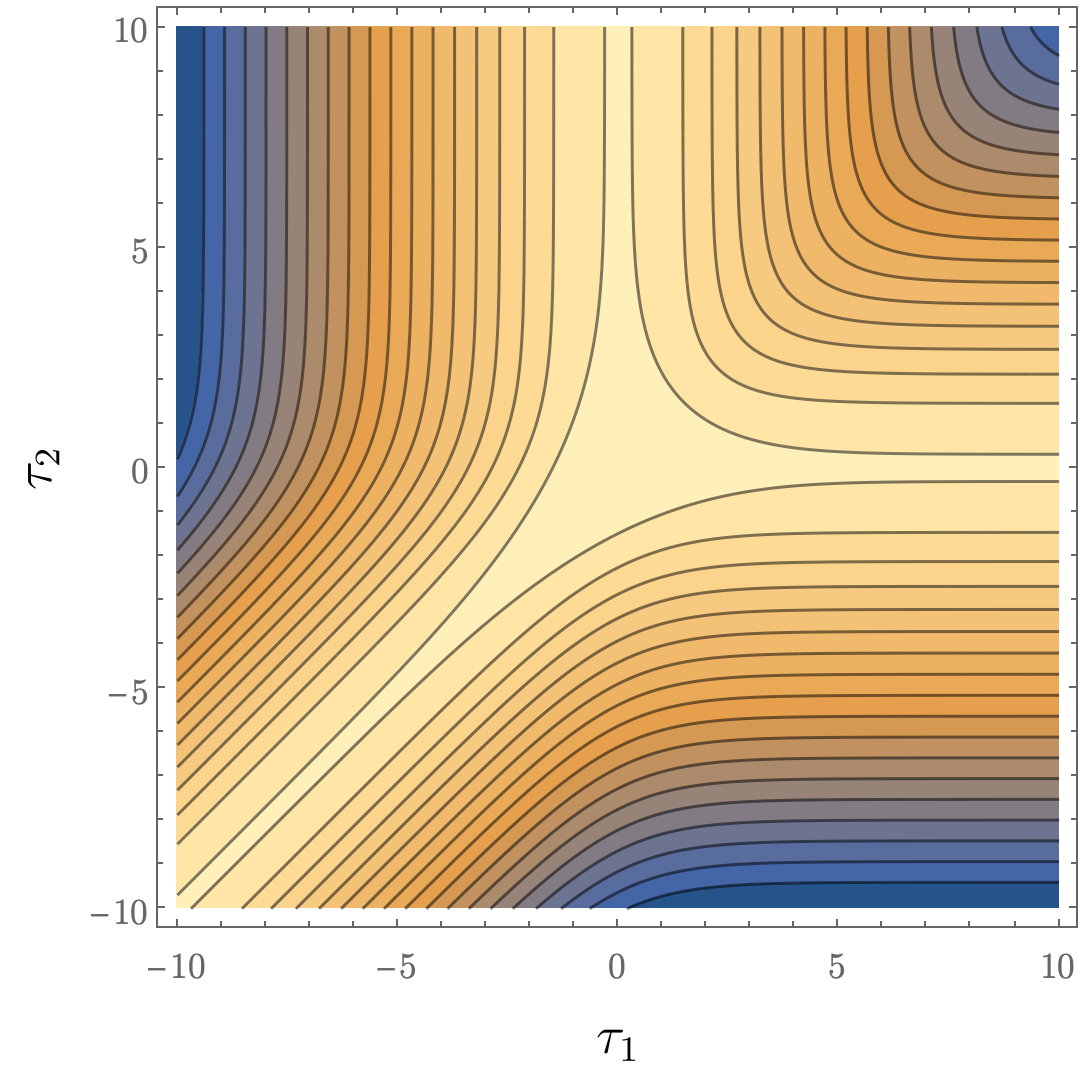}
    \includegraphics[scale=0.71]{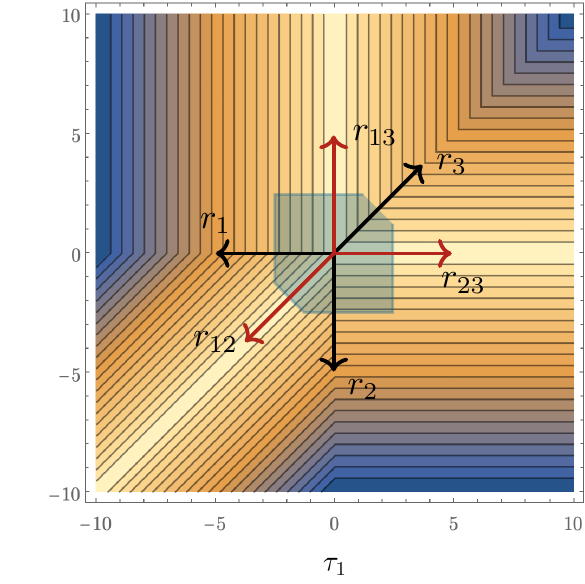}
    \hfill
    \caption{Left: Logarithm of the integrand of \eqref{eq:sunrise} in the $(\tau_1, \tau_2)$-plane with $\tau_3=0$. Right: Tropical approximation with the Trop function and the corresponding $\SS_{\mathrm{sun}}$ polytope with $c>0$ overlayed in blue. The three divergent directions (more yellow) are the three rays with $d_{\gamma}=0$ (red), giving logarithmic UV divergences associated to bubble subdiagrams.}
    \label{fig:sunrise}
\end{figure}

{\bf Massive sunrise.} Let us now have a look at a simple example of UV divergences. We consider the sunrise diagram with massive internal legs, which according to \eqref{eq:I} gives
\begin{align}\label{eq:sunrise}
\I_{\mathrm{sun}} &= \Gamma(3{-}\D) \int \frac{\d^3 \alpha}{\mathrm{GL}(1)}\\ &\frac{\U_{\mathrm{sun}}^{3-3\D/2}}{(p_1^2 \alpha_1 \alpha_2 \alpha_3 - (m_1^2\alpha_1+ m_2^2\alpha_2+ m_3^2\alpha_3)\U_{\mathrm{sun}})^{3-\D}},\nn
\end{align}
where
\begin{equation}
\U_{\mathrm{sun}} = \alpha_1 \alpha_2 + \alpha_2\alpha_3 + \alpha_3 \alpha_1.
\end{equation}
Setting $\alpha_3 = e^{\tau_3}=1$, the Trop function in $\D=4{-}2\varepsilon$ reads
\begin{align}\label{eq:sunrise-Trop}
&\Trop = \tau_1 {+} \tau_2 - (3{-}3\varepsilon) \max(\tau_1 {+} \tau_2,\, \tau_2,\, \tau_1)\\
&+ (1{-}2\varepsilon) \max(\tau_1 {+} \tau_2,\,  2\tau_1 {+} \tau_2,\, \tau_1 {+} 2\tau_2,\,  2\tau_2,\, \tau_2, \,\tau_1,\, 2\tau_1).\nn
\end{align}
It has six domains of linearity divided by six rays, which are plotted in Fig.~\ref{fig:sunrise} together with the integrand it is approximating.
One can immediately identify three divergent rays, along which $\Trop = {\cal O}(\varepsilon)$. Let us focus on the one pointing to the bottom-left, which comes from rescaling the Schwinger parameters according to
\begin{equation}
(\alpha_1 : \alpha_2 : \alpha_3) = ( \tfrac{1}{\rho} : \tfrac{1}{\rho} : 1)
\end{equation}
with $\rho \to \infty$. It is therefore the UV divergence coming from shrinking the loop $\gamma_{12}$, in agreement with \eqref{eq:Trop-gamma}. In order to compute the divergence, we identify the divergent ray $r_{12}$ (dual to a one-dimensional facet $F_{12}$) and its orthogonal direction:
\begin{equation}
w_1^{(F_{12})} = \begin{pmatrix}
    -1 \\ -1
    \end{pmatrix},
    \qquad
    \bar{w}_1^{(F_{12})}  = \begin{pmatrix}
    1 \\ 0
    \end{pmatrix}.
\end{equation}
Using \eqref{eq:volC} in $\D=4{-}2\varepsilon$, the volume of this one-dimensional cone is
\begin{equation}
\mathrm{vol}(C_{F_{12}}) = \frac{1}{d_{12}} = \frac{1}{\varepsilon}.
\end{equation}
The coefficient of this divergence is the leftover one-dimensional integral, where we drop subleading monomials from the Symanzik polynomials according to
\begin{equation}
\U_{\mathrm{sun}} \to \alpha_3(\alpha_1 + \alpha_2), \quad \F_{\mathrm{sun}} \to -m_3^2 \alpha_3 \U_{\mathrm{sun}},
\end{equation}
followed by setting $\alpha_2 = \alpha_3 = 1$ to fix the $\mathrm{GL}(1)^2$ symmetry. This gives:
\begin{equation}
\J_{C_{F_{12}}} = -m_3^2 \int_{\mathbb{R}_+} \frac{\d \hat{\alpha}_1}{(1+\hat{\alpha}_1)^2} = -m_3^2.
\end{equation}

By symmetry, the remaining two divergent rays give bubble UV singularities for shrinking $\gamma_{23}$ and $\gamma_{31}$ and the total leading divergence is their sum.
Note that in this case, the sunrise also has an overall power-law divergence coming from the prefactor in \eqref{eq:sunrise}, $\Gamma(-1+2\varepsilon) = -\frac{1}{2\varepsilon} + \ldots$.
Taking this into account, at the leading order we have
\begin{equation}
\I_{\mathrm{sun}} = \frac{1}{2\varepsilon^2} \left( m_1^2 + m_2^2 + m_3^2 \right) + \ldots
\end{equation}
in agreement with \cite{Adams:2015gva}.

\begin{figure}[!t]
    \centering
    \includegraphics[]{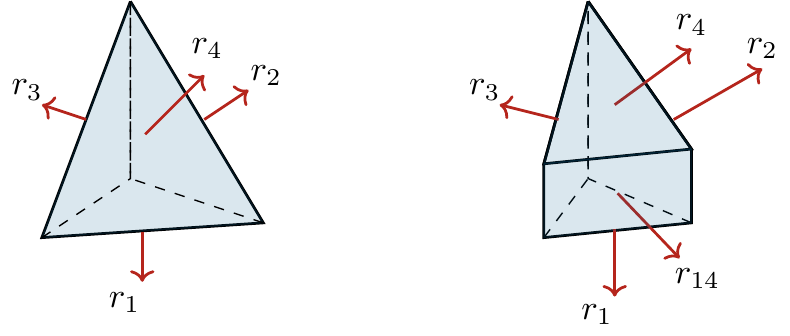}
    \caption{Left: The $\SS$ polytope for the box with non-zero corner masses. Right: A new facet associated with the IR subgraph $\gamma_{14}$ emerges in the three-mass box case. This kind of facet corresponds to inequalities of the second type in \eqref{fdefn}.}
    \label{fig:twopolytopes}
\end{figure}

\par{\bf Three-mass box.} Let us return back to the example from Sec.~\ref{sec:IR} illustrated in Fig.~\ref{IRcorner}. In $\D=4{-}2\varepsilon$, the massless box with three massive corners has a single logarithmic divergence with a non-trivial finite coefficient. The Feynman integral \eqref{eq:I} in this case is 
\begin{align}
   \I_{\mathrm{box}} &= \Gamma(4-\tfrac{D}{2}) \int \frac{\d^4\alpha}{\text{GL}(1)}\\
   &\!\!\!\frac{(\alpha_1+\alpha_2+\alpha_3+\alpha_4)^{4-\D}}{(s \alpha_1\alpha_3+t\alpha_2\alpha_4+p_1^2\alpha_1\alpha_2+p_3^2\alpha_3\alpha_4+p_4^2\alpha_1\alpha_4)^{4-\D/2}},\nn
\end{align}
where $p_2^2 = 0$ is the only massless corner. 
The $\SS$ polytope is displayed in Fig.~\ref{fig:twopolytopes}.
There is a single divergent ray $r_{14}$ corresponding to subgraph $\gamma_{14}$ and the value of $\Trop$ along this ray is $d_{G/\gamma_{14}} = 2-\D/2$, which is indeed divergent in $\D \leq 4$. This is the collinear divergence of the two internal propagators adjacent to the massless corner, parametrized by
\begin{equation}
(\alpha_1 : \alpha_2 : \alpha_3 : \alpha_4) = ( 1 : \rho : \rho : 1 )
\end{equation}
with $\rho \to \infty$.
It is dual to the facet $F_{14}$ with normal $r_{14}$ in Fig.~\ref{fig:twopolytopes}. According to \eqref{eq:volC}, we have 
\begin{equation}
\mathrm{vol}(C_{F_{14}}) = -\frac{1}{d_{G/\gamma_{14}}} = -\frac{1}{\varepsilon}.
\end{equation}
In this limit, we drop subleading monomials from Symanzik polynomials and are left with
\begin{align}
\U_{\mathrm{box}} &\to \alpha_2 + \alpha_3,\\
\F_{\mathrm{box}} &\to s \alpha_1\alpha_3+t\alpha_2\alpha_4+p_1^2\alpha_1\alpha_2+p_3^2\alpha_3\alpha_4.
\end{align}
This gives the kinematic coefficient of the divergence:
\begin{equation}
\J_{C_{F_{14}}} = \int_{\mathbb{R}^2_+} \frac{\d\hat \alpha_1 \d\hat \alpha_2 }{(s \hat \alpha_1+t\hat\alpha_2+p_1^2\hat\alpha_1\hat\alpha_2+p_3^2)^2},
\end{equation}
where the choice of variables yielding unit  Jacobian is equivalent to setting $\alpha_3 \to 1$ for the overall $\mathrm{GL}(1)$ and $\alpha_4 \to 1$ for the enhanced $\mathrm{GL}(1)$ symmetry along the divergent ray, yielding the finite integral above. In summary, taking into account the overall $\Gamma(4-\tfrac{\D}{2})$, the leading divergence contributes
\begin{equation}
    \I_{\mathrm{box}} = -\frac{1}{\varepsilon} \frac{\log\left(\frac{p_1^2 p_3^2}{s t} \right)}{st-p_1^2 p_3^2} + \ldots
\end{equation}
in agreement with \cite{Bern:1993kr}.

\bibliographystyle{JHEP}
\bibliography{references}

\providecommand{\href}[2]{#2}\begingroup\raggedright\begin{thebibliography}{10}

\bibitem{10.1007/BF02392399}
N.~N. Bogoliubow and O.~S. Parasiuk, \emph{{Über die Multiplikation der
  Kausalfunktionen in der Quantentheorie der Felder}},
  \href{http://dx.doi.org/10.1007/BF02392399}{\emph{Acta Mathematica}
  {\bfseries 97} (1957) 227 -- 266}.

\bibitem{Hepp1966}
K.~Hepp, \emph{{Proof of the Bogoliubov-Parasiuk theorem on renormalization}},
  \href{http://dx.doi.org/10.1007/BF01773358}{\emph{Communications in
  Mathematical Physics} {\bfseries 2} (Dec, 1966) 301--326}.

\bibitem{Zimmermann1969}
W.~Zimmermann, \emph{{Convergence of Bogoliubov's method of renormalization in
  momentum space}},
  \href{http://dx.doi.org/10.1007/BF01645676}{\emph{Communications in
  Mathematical Physics} {\bfseries 15} (Sep, 1969) 208--234}.

\bibitem{Connes:1998qv}
A.~Connes and D.~Kreimer, \emph{{Hopf algebras, renormalization and
  noncommutative geometry}},
  \href{http://dx.doi.org/10.1007/s002200050499}{\emph{Commun. Math. Phys.}
  {\bfseries 199} (1998) 203--242},
  [\href{https://arxiv.org/abs/hep-th/9808042}{{\ttfamily hep-th/9808042}}].

\bibitem{Connes:1999yr}
A.~Connes and D.~Kreimer, \emph{{Renormalization in quantum field theory and
  the Riemann-Hilbert problem. 1. The Hopf algebra structure of graphs and the
  main theorem}}, \href{http://dx.doi.org/10.1007/s002200050779}{\emph{Commun.
  Math. Phys.} {\bfseries 210} (2000) 249--273},
  [\href{https://arxiv.org/abs/hep-th/9912092}{{\ttfamily hep-th/9912092}}].

\bibitem{Connes:2000fe}
A.~Connes and D.~Kreimer, \emph{{Renormalization in quantum field theory and
  the Riemann-Hilbert problem. 2. The beta function, diffeomorphisms and the
  renormalization group}},
  \href{http://dx.doi.org/10.1007/PL00005547}{\emph{Commun. Math. Phys.}
  {\bfseries 216} (2001) 215--241},
  [\href{https://arxiv.org/abs/hep-th/0003188}{{\ttfamily hep-th/0003188}}].

\bibitem{Mueller:1979ih}
A.~H. Mueller, \emph{{On the Asymptotic Behavior of the Sudakov Form-factor}},
  \href{http://dx.doi.org/10.1103/PhysRevD.20.2037}{\emph{Phys. Rev. D}
  {\bfseries 20} (1979) 2037}.

\bibitem{Collins:1980ih}
J.~C. Collins, \emph{{Algorithm to Compute Corrections to the Sudakov
  Form-factor}}, \href{http://dx.doi.org/10.1103/PhysRevD.22.1478}{\emph{Phys.
  Rev. D} {\bfseries 22} (1980) 1478}.

\bibitem{Sen:1981sd}
A.~Sen, \emph{{Asymptotic Behavior of the Sudakov Form-Factor in QCD}},
  \href{http://dx.doi.org/10.1103/PhysRevD.24.3281}{\emph{Phys. Rev. D}
  {\bfseries 24} (1981) 3281}.

\bibitem{Korchemsky:1988si}
G.~P. Korchemsky, \emph{{Asymptotics of the Altarelli-Parisi-Lipatov Evolution
  Kernels of Parton Distributions}},
  \href{http://dx.doi.org/10.1142/S0217732389001453}{\emph{Mod. Phys. Lett. A}
  {\bfseries 4} (1989) 1257--1276}.

\bibitem{Sterman:2002qn}
G.~F. Sterman and M.~E. Tejeda-Yeomans, \emph{{Multiloop amplitudes and
  resummation}},
  \href{http://dx.doi.org/10.1016/S0370-2693(02)03100-3}{\emph{Phys. Lett. B}
  {\bfseries 552} (2003) 48--56},
  [\href{https://arxiv.org/abs/hep-ph/0210130}{{\ttfamily hep-ph/0210130}}].

\bibitem{Gardi:2009qi}
E.~Gardi and L.~Magnea, \emph{{Factorization constraints for soft anomalous
  dimensions in QCD scattering amplitudes}},
  \href{http://dx.doi.org/10.1088/1126-6708/2009/03/079}{\emph{JHEP} {\bfseries
  03} (2009) 079}, [\href{https://arxiv.org/abs/0901.1091}{{\ttfamily
  0901.1091}}].

\bibitem{Agarwal:2021ais}
N.~Agarwal, L.~Magnea, C.~Signorile-Signorile and A.~Tripathi, \emph{{The
  Infrared Structure of Perturbative Gauge Theories}},
  \href{https://arxiv.org/abs/2112.07099}{{\ttfamily 2112.07099}}.

\bibitem{Collins:1989gx}
J.~C. Collins, D.~E. Soper and G.~F. Sterman, \emph{{Factorization of Hard
  Processes in QCD}},
  \href{http://dx.doi.org/10.1142/9789814503266_0001}{\emph{Adv. Ser. Direct.
  High Energy Phys.} {\bfseries 5} (1989) 1--91},
  [\href{https://arxiv.org/abs/hep-ph/0409313}{{\ttfamily hep-ph/0409313}}].

\bibitem{Becher:2009qa}
T.~Becher and M.~Neubert, \emph{{On the Structure of Infrared Singularities of
  Gauge-Theory Amplitudes}},
  \href{http://dx.doi.org/10.1088/1126-6708/2009/06/081}{\emph{JHEP} {\bfseries
  06} (2009) 081}, [\href{https://arxiv.org/abs/0903.1126}{{\ttfamily
  0903.1126}}].

\bibitem{StewartLectures}
{Iain W. Stewart}, \emph{{Lectures on the Soft-Collinear Effective Theory}},
  {\emph{{\href{https://ocw.mit.edu/courses/physics/8-851-effective-field-theory-spring-2013/lecture-notes/MIT8_851S13_scetnotes.pdf}{EFT
  Course 8.851, SCET Lecture Notes, Massachusetts Institute of Technology}}}
  (2013) }.

\bibitem{Becher:2014oda}
T.~Becher, A.~Broggio and A.~Ferroglia, \emph{{Introduction to Soft-Collinear
  Effective Theory}}, vol.~896.
\newblock Springer, 2015,
  \href{http://dx.doi.org/10.1007/978-3-319-14848-9}{10.1007/978-3-319-14848-9}.

\bibitem{Ma:2019hjq}
Y.~Ma, \emph{{A Forest Formula to Subtract Infrared Singularities in Amplitudes
  for Wide-angle Scattering}},
  \href{http://dx.doi.org/10.1007/JHEP05(2020)012}{\emph{JHEP} {\bfseries 05}
  (2020) 012}, [\href{https://arxiv.org/abs/1910.11304}{{\ttfamily
  1910.11304}}].

\bibitem{Beekveldt:2020kzk}
R.~Beekveldt, M.~Borinsky and F.~Herzog, \emph{{The Hopf algebra structure of
  the R$^{*}$-operation}},
  \href{http://dx.doi.org/10.1007/JHEP07(2020)061}{\emph{JHEP} {\bfseries 07}
  (2020) 061}, [\href{https://arxiv.org/abs/2003.04301}{{\ttfamily
  2003.04301}}].

\bibitem{maclagan2015introduction}
D.~Maclagan and B.~Sturmfels, \emph{Introduction to Tropical Geometry}.
\newblock Graduate Studies in Mathematics. American Mathematical Society, 2015.

\bibitem{brown2011multiple}
F.~Brown, \emph{{Multiple zeta values and periods: from moduli spaces to
  Feynman integrals}}, {\emph{Contemp. Math} (2011) 27--52}.

\bibitem{Brown:2015fyf}
F.~Brown, \emph{{Feynman amplitudes, coaction principle, and cosmic Galois
  group}}, \href{http://dx.doi.org/10.4310/CNTP.2017.v11.n3.a1}{\emph{Commun.
  Num. Theor. Phys.} {\bfseries 11} (2017) 453--556},
  [\href{https://arxiv.org/abs/1512.06409}{{\ttfamily 1512.06409}}].

\bibitem{Schultka:2018nrs}
K.~Schultka, \emph{{Toric geometry and regularization of Feynman integrals}},
  \href{https://arxiv.org/abs/1806.01086}{{\ttfamily 1806.01086}}.

\bibitem{Borinsky:2020rqs}
M.~Borinsky, \emph{{Tropical Monte Carlo quadrature for Feynman integrals}},
  \href{https://arxiv.org/abs/2008.12310}{{\ttfamily 2008.12310}}.

\bibitem{Mizera:2021icv}
S.~Mizera and S.~Telen, \emph{{Landau Discriminants}},
  \href{https://arxiv.org/abs/2109.08036}{{\ttfamily 2109.08036}}.

\bibitem{Binoth:2000ps}
T.~Binoth and G.~Heinrich, \emph{{An automatized algorithm to compute infrared
  divergent multiloop integrals}},
  \href{http://dx.doi.org/10.1016/S0550-3213(00)00429-6}{\emph{Nucl. Phys. B}
  {\bfseries 585} (2000) 741--759},
  [\href{https://arxiv.org/abs/hep-ph/0004013}{{\ttfamily hep-ph/0004013}}].

\bibitem{Binoth:2003ak}
T.~Binoth and G.~Heinrich, \emph{{Numerical evaluation of multiloop integrals
  by sector decomposition}},
  \href{http://dx.doi.org/10.1016/j.nuclphysb.2003.12.023}{\emph{Nucl. Phys. B}
  {\bfseries 680} (2004) 375--388},
  [\href{https://arxiv.org/abs/hep-ph/0305234}{{\ttfamily hep-ph/0305234}}].

\bibitem{Bogner:2007cr}
C.~Bogner and S.~Weinzierl, \emph{{Resolution of singularities for multi-loop
  integrals}}, \href{http://dx.doi.org/10.1016/j.cpc.2007.11.012}{\emph{Comput.
  Phys. Commun.} {\bfseries 178} (2008) 596--610},
  [\href{https://arxiv.org/abs/0709.4092}{{\ttfamily 0709.4092}}].

\bibitem{Kaneko:2009qx}
T.~Kaneko and T.~Ueda, \emph{{A Geometric method of sector decomposition}},
  \href{http://dx.doi.org/10.1016/j.cpc.2010.04.001}{\emph{Comput. Phys.
  Commun.} {\bfseries 181} (2010) 1352--1361},
  [\href{https://arxiv.org/abs/0908.2897}{{\ttfamily 0908.2897}}].

\bibitem{Kaneko:2010kj}
T.~Kaneko and T.~Ueda, \emph{{Sector Decomposition Via Computational
  Geometry}}, \href{http://dx.doi.org/10.22323/1.093.0082}{\emph{PoS}
  {\bfseries ACAT2010} (2010) 082},
  [\href{https://arxiv.org/abs/1004.5490}{{\ttfamily 1004.5490}}].

\bibitem{Borowka:2015mxa}
S.~Borowka, G.~Heinrich, S.~P. Jones, M.~Kerner, J.~Schlenk and T.~Zirke,
  \emph{{SecDec-3.0: numerical evaluation of multi-scale integrals beyond one
  loop}}, \href{http://dx.doi.org/10.1016/j.cpc.2015.05.022}{\emph{Comput.
  Phys. Commun.} {\bfseries 196} (2015) 470--491},
  [\href{https://arxiv.org/abs/1502.06595}{{\ttfamily 1502.06595}}].

\bibitem{Schlenk:2016epj}
J.~K. Schlenk, \emph{{Techniques for higher order corrections and their
  application to LHC phenomenology}}, Ph.D. thesis, Munich, Tech. U., 8, 2016.

\bibitem{Heinrich:2021dbf}
G.~Heinrich, S.~Jahn, S.~P. Jones, M.~Kerner, F.~Langer, V.~Magerya,
  A.~P\"oldaru et~al., \emph{{Expansion by regions with pySecDec}},
  \href{http://dx.doi.org/10.1016/j.cpc.2021.108267}{\emph{Comput. Phys.
  Commun.} {\bfseries 273} (2022) 108267},
  [\href{https://arxiv.org/abs/2108.10807}{{\ttfamily 2108.10807}}].

\bibitem{Pak:2010pt}
A.~Pak and A.~Smirnov, \emph{{Geometric approach to asymptotic expansion of
  Feynman integrals}},
  \href{http://dx.doi.org/10.1140/epjc/s10052-011-1626-1}{\emph{Eur. Phys. J.
  C} {\bfseries 71} (2011) 1626},
  [\href{https://arxiv.org/abs/1011.4863}{{\ttfamily 1011.4863}}].

\bibitem{Ananthanarayan:2018tog}
B.~Ananthanarayan, A.~Pal, S.~Ramanan and R.~Sarkar, \emph{{Unveiling Regions
  in multi-scale Feynman Integrals using Singularities and Power Geometry}},
  \href{http://dx.doi.org/10.1140/epjc/s10052-019-6533-x}{\emph{Eur. Phys. J.
  C} {\bfseries 79} (2019) 57},
  [\href{https://arxiv.org/abs/1810.06270}{{\ttfamily 1810.06270}}].

\bibitem{Semenova:2018cwy}
T.~Y. Semenova, A.~V. Smirnov and V.~A. Smirnov, \emph{{On the status of
  expansion by regions}},
  \href{http://dx.doi.org/10.1140/epjc/s10052-019-6653-3}{\emph{Eur. Phys. J.
  C} {\bfseries 79} (2019) 136},
  [\href{https://arxiv.org/abs/1809.04325}{{\ttfamily 1809.04325}}].

\bibitem{Ananthanarayan:2020fhl}
B.~Ananthanarayan, S.~Banik, S.~Friot and S.~Ghosh, \emph{{Multiple Series
  Representations of N-fold Mellin-Barnes Integrals}},
  \href{http://dx.doi.org/10.1103/PhysRevLett.127.151601}{\emph{Phys. Rev.
  Lett.} {\bfseries 127} (2021) 151601},
  [\href{https://arxiv.org/abs/2012.15108}{{\ttfamily 2012.15108}}].

\bibitem{Tellander:2021xdz}
F.~Tellander and M.~Helmer, \emph{{Cohen-Macaulay Property of Feynman
  Integrals}},  \href{https://arxiv.org/abs/2108.01410}{{\ttfamily
  2108.01410}}.

\bibitem{Bloch2006}
S.~Bloch, H.~Esnault and D.~Kreimer, \emph{{On Motives Associated to Graph
  Polynomials}},
  \href{http://dx.doi.org/10.1007/s00220-006-0040-2}{\emph{Commun. Math. Phys.}
  {\bfseries 267} (Oct, 2006) 181--225}.

\bibitem{Tourkine:2013rda}
P.~Tourkine, \emph{{Tropical Amplitudes}},
  \href{http://dx.doi.org/10.1007/s00023-017-0560-7}{\emph{Annales Henri
  Poincare} {\bfseries 18} (2017) 2199--2249},
  [\href{https://arxiv.org/abs/1309.3551}{{\ttfamily 1309.3551}}].

\bibitem{Panzer:2019yxl}
E.~Panzer, \emph{{Hepp's bound for Feynman graphs and matroids}},
  \href{https://arxiv.org/abs/1908.09820}{{\ttfamily 1908.09820}}.

\bibitem{Cachazo:2019ngv}
F.~Cachazo, N.~Early, A.~Guevara and S.~Mizera, \emph{{Scattering Equations:
  From Projective Spaces to Tropical Grassmannians}},
  \href{http://dx.doi.org/10.1007/JHEP06(2019)039}{\emph{JHEP} {\bfseries 06}
  (2019) 039}, [\href{https://arxiv.org/abs/1903.08904}{{\ttfamily
  1903.08904}}].

\bibitem{Drummond:2020kqg}
J.~Drummond, J.~Foster, O.~G\"urdo\u{g}an and C.~Kalousios, \emph{{Tropical
  fans, scattering equations and amplitudes}},
  \href{https://arxiv.org/abs/2002.04624}{{\ttfamily 2002.04624}}.

\bibitem{Arkani-Hamed:2020cig}
N.~Arkani-Hamed, T.~Lam and M.~Spradlin, \emph{{Positive configuration space}},
  \href{http://dx.doi.org/10.1007/s00220-021-04041-x}{\emph{Commun. Math.
  Phys.} {\bfseries 384} (2021) 909--954},
  [\href{https://arxiv.org/abs/2003.03904}{{\ttfamily 2003.03904}}].

\bibitem{Drummond:2019cxm}
J.~Drummond, J.~Foster, O.~G\"urdogan and C.~Kalousios, \emph{{Algebraic
  singularities of scattering amplitudes from tropical geometry}},
  \href{http://dx.doi.org/10.1007/JHEP04(2021)002}{\emph{JHEP} {\bfseries 04}
  (2021) 002}, [\href{https://arxiv.org/abs/1912.08217}{{\ttfamily
  1912.08217}}].

\bibitem{Arkani-Hamed:2019mrd}
N.~Arkani-Hamed, S.~He and T.~Lam, \emph{{Stringy canonical forms}},
  \href{http://dx.doi.org/10.1007/JHEP02(2021)069}{\emph{JHEP} {\bfseries 02}
  (2021) 069}, [\href{https://arxiv.org/abs/1912.08707}{{\ttfamily
  1912.08707}}].

\bibitem{Arkani-Hamed:2019plo}
N.~Arkani-Hamed, S.~He, T.~Lam and H.~Thomas, \emph{{Binary Geometries,
  Generalized Particles and Strings, and Cluster Algebras}},
  \href{https://arxiv.org/abs/1912.11764}{{\ttfamily 1912.11764}}.

\bibitem{Arkani-Hamed:2020tuz}
N.~Arkani-Hamed, S.~He and T.~Lam, \emph{{Cluster configuration spaces of
  finite type}},  \href{https://arxiv.org/abs/2005.11419}{{\ttfamily
  2005.11419}}.

\bibitem{Chicherin:2020umh}
D.~Chicherin, J.~M. Henn and G.~Papathanasiou, \emph{{Cluster algebras for
  Feynman integrals}},
  \href{http://dx.doi.org/10.1103/PhysRevLett.126.091603}{\emph{Phys. Rev.
  Lett.} {\bfseries 126} (2021) 091603},
  [\href{https://arxiv.org/abs/2012.12285}{{\ttfamily 2012.12285}}].

\bibitem{Cachazo:2020wgu}
F.~Cachazo and N.~Early, \emph{{Planar Kinematics: Cyclic Fixed Points, Mirror
  Superpotential, k-Dimensional Catalan Numbers, and Root Polytopes}},
  \href{https://arxiv.org/abs/2010.09708}{{\ttfamily 2010.09708}}.

\bibitem{He:2021non}
S.~He, Z.~Li and Q.~Yang, \emph{{Truncated cluster algebras and Feynman
  integrals with algebraic letters}},
  \href{https://arxiv.org/abs/2106.09314}{{\ttfamily 2106.09314}}.

\bibitem{Mastrolia:2018uzb}
P.~Mastrolia and S.~Mizera, \emph{{Feynman Integrals and Intersection Theory}},
  \href{http://dx.doi.org/10.1007/JHEP02(2019)139}{\emph{JHEP} {\bfseries 02}
  (2019) 139}, [\href{https://arxiv.org/abs/1810.03818}{{\ttfamily
  1810.03818}}].

\bibitem{delaCruz:2019skx}
L.~de~la Cruz, \emph{{Feynman integrals as A-hypergeometric functions}},
  \href{http://dx.doi.org/10.1007/JHEP12(2019)123}{\emph{JHEP} {\bfseries 12}
  (2019) 123}, [\href{https://arxiv.org/abs/1907.00507}{{\ttfamily
  1907.00507}}].

\bibitem{Klausen:2019hrg}
R.~P. Klausen, \emph{{Hypergeometric Series Representations of Feynman
  Integrals by GKZ Hypergeometric Systems}},
  \href{http://dx.doi.org/10.1007/JHEP04(2020)121}{\emph{JHEP} {\bfseries 04}
  (2020) 121}, [\href{https://arxiv.org/abs/1910.08651}{{\ttfamily
  1910.08651}}].

\bibitem{Feng:2019bdx}
T.-F. Feng, C.-H. Chang, J.-B. Chen and H.-B. Zhang, \emph{{GKZ-hypergeometric
  systems for Feynman integrals}},
  \href{http://dx.doi.org/10.1016/j.nuclphysb.2020.114952}{\emph{Nucl. Phys. B}
  {\bfseries 953} (2020) 114952},
  [\href{https://arxiv.org/abs/1912.01726}{{\ttfamily 1912.01726}}].

\bibitem{Abreu:2019wzk}
S.~Abreu, R.~Britto, C.~Duhr, E.~Gardi and J.~Matthew, \emph{{From positive
  geometries to a coaction on hypergeometric functions}},
  \href{http://dx.doi.org/10.1007/JHEP02(2020)122}{\emph{JHEP} {\bfseries 02}
  (2020) 122}, [\href{https://arxiv.org/abs/1910.08358}{{\ttfamily
  1910.08358}}].

\bibitem{Brown:2011pj}
F.~Brown and D.~Kreimer, \emph{{Angles, Scales and Parametric
  Renormalization}},
  \href{http://dx.doi.org/10.1007/s11005-013-0625-6}{\emph{Lett. Math. Phys.}
  {\bfseries 103} (2013) 933--1007},
  [\href{https://arxiv.org/abs/1112.1180}{{\ttfamily 1112.1180}}].

\bibitem{AHHM}
N.~Arkani-Hamed, A.~Hillman and S.~Mizera, \emph{in preparation}, .

\bibitem{nakanishi1971graph}
N.~Nakanishi, \emph{{Graph Theory and Feynman Integrals}}.
\newblock Mathematics and its applications: a series of monographs and texts.
  Gordon and Breach, 1971.

\bibitem{gelfand2009discriminants}
I.~Gelfand, M.~Kapranov and A.~Zelevinsky, \emph{Discriminants, Resultants, and
  Multidimensional Determinants}.
\newblock Modern Birkh{\"a}user Classics. Birkh{\"a}user Boston, 2009,
  \href{http://dx.doi.org/10.1007/978-0-8176-4771-1}{10.1007/978-0-8176-4771-1}.

\bibitem{Nilsson2013}
L.~Nilsson and M.~Passare, \emph{Mellin transforms of multivariate rational
  functions}, \href{http://dx.doi.org/10.1007/s12220-011-9235-7}{\emph{Journal
  of Geometric Analysis} {\bfseries 23} (Jan, 2013) 24--46},
  [\href{https://arxiv.org/abs/1010.5060}{{\ttfamily 1010.5060}}].

\bibitem{berkesch2013eulermellin}
C.~Berkesch, J.~Forsgård and M.~Passare, \emph{{Euler--Mellin integrals and
  A-hypergeometric functions}},
  \href{https://arxiv.org/abs/1103.6273}{{\ttfamily 1103.6273}}.

\bibitem{PhysRev.118.838}
S.~Weinberg, \emph{{High-Energy Behavior in Quantum Field Theory}},
  \href{http://dx.doi.org/10.1103/PhysRev.118.838}{\emph{Phys. Rev.} {\bfseries
  118} (May, 1960) 838--849}.

\bibitem{Lowenstein:1975rg}
J.~H. Lowenstein and W.~Zimmermann, \emph{{The Power Counting Theorem for
  Feynman Integrals with Massless Propagators}},
  \href{http://dx.doi.org/10.1007/BF01609059}{\emph{Commun. Math. Phys.}
  {\bfseries 44} (1975) 73--86}.

\bibitem{YelleshpurSrikant:2019khx}
A.~Yelleshpur~Srikant, \emph{{Spherical Contours, IR Divergences and the
  geometry of Feynman parameter integrands at one loop}},
  \href{http://dx.doi.org/10.1007/JHEP07(2020)236}{\emph{JHEP} {\bfseries 07}
  (2020) 236}, [\href{https://arxiv.org/abs/1907.05429}{{\ttfamily
  1907.05429}}].

\bibitem{Gurdogan:2015csr}
O.~G\"urdo\u{g}an and V.~Kazakov, \emph{{New Integrable 4D Quantum Field
  Theories from Strongly Deformed Planar $\mathcal N = $ 4 Supersymmetric
  Yang-Mills Theory}},
  \href{http://dx.doi.org/10.1103/PhysRevLett.117.201602}{\emph{Phys. Rev.
  Lett.} {\bfseries 117} (2016) 201602},
  [\href{https://arxiv.org/abs/1512.06704}{{\ttfamily 1512.06704}}].

\bibitem{Chicherin:2017frs}
D.~Chicherin, V.~Kazakov, F.~Loebbert, D.~M\"uller and D.-l. Zhong,
  \emph{{Yangian Symmetry for Fishnet Feynman Graphs}},
  \href{http://dx.doi.org/10.1103/PhysRevD.96.121901}{\emph{Phys. Rev. D}
  {\bfseries 96} (2017) 121901},
  [\href{https://arxiv.org/abs/1708.00007}{{\ttfamily 1708.00007}}].

\bibitem{Loebbert:2020tje}
F.~Loebbert and J.~Miczajka, \emph{{Massive Fishnets}},
  \href{http://dx.doi.org/10.1007/JHEP12(2020)197}{\emph{JHEP} {\bfseries 12}
  (2020) 197}, [\href{https://arxiv.org/abs/2008.11739}{{\ttfamily
  2008.11739}}].

\bibitem{MultiCatalan}
{OEIS Foundation Inc.}, \emph{{Multidimensional Catalan numbers, Entry A060854
  in The On-Line Encyclopedia of Integer Sequences}},
  {\emph{{\normalfont[\url{https://oeis.org/A060854}]}} (2022) }.

\bibitem{Basso:2021omx}
B.~Basso, L.~J. Dixon, D.~A. Kosower, A.~Krajenbrink and D.-l. Zhong,
  \emph{{Fishnet four-point integrals: integrable representations and
  thermodynamic limits}},
  \href{http://dx.doi.org/10.1007/JHEP07(2021)168}{\emph{JHEP} {\bfseries 07}
  (2021) 168}, [\href{https://arxiv.org/abs/2105.10514}{{\ttfamily
  2105.10514}}].

\bibitem{Olivucci:2021pss}
E.~Olivucci and P.~Vieira, \emph{{Stampedes I: Fishnet OPE and Octagon
  Bootstrap with Nonzero Bridges}},
  \href{https://arxiv.org/abs/2111.12131}{{\ttfamily 2111.12131}}.

\bibitem{BrownIHES}
F.~Brown, ``{Motivic periods and the cosmic Galois group (IHES, May 2015)}.''

\bibitem{Adams:2015gva}
L.~Adams, C.~Bogner and S.~Weinzierl, \emph{{The two-loop sunrise integral
  around four space-time dimensions and generalisations of the Clausen and
  Glaisher functions towards the elliptic case}},
  \href{http://dx.doi.org/10.1063/1.4926985}{\emph{J. Math. Phys.} {\bfseries
  56} (2015) 072303}, [\href{https://arxiv.org/abs/1504.03255}{{\ttfamily
  1504.03255}}].

\bibitem{Bern:1993kr}
Z.~Bern, L.~J. Dixon and D.~A. Kosower, \emph{{Dimensionally regulated pentagon
  integrals}},
  \href{http://dx.doi.org/10.1016/0550-3213(94)90398-0}{\emph{Nucl. Phys. B}
  {\bfseries 412} (1994) 751--816},
  [\href{https://arxiv.org/abs/hep-ph/9306240}{{\ttfamily hep-ph/9306240}}].

\end{thebibliography}\endgroup

\end{document}